# The secondary classification of unequilibrated chondrites

Emmanuel Jacquet[1], Béatrice Doisneau[1]

[1]Institut de Minéralogie, de Physique des Matériaux et de Cosmochimie (IMPMC), Muséum national d'Histoire naturelle, Sorbonne Université, CNRS; CP52, 57 rue Cuvier, 75005 Paris, France.

## Abstract

The multiplication of decimal petrologic schemes for different or the same chondrite groups evinces a lack of unified guiding principle in the secondary classification of type 1-3 chondrites. We show that the current OC, R and CO classifications can be *a posteriori* unified, with only minor reclassifications, if the decimal part of the subtype is defined as the ratio $m=Fa_I/Fa_{II}$ of the mean fayalite contents of type I and type II chondrules rounded to the nearest tenth (with adaptations from Cr systematics for the lowest subtypes following Grossman and Brearley 2005). This parameter is more efficiently evaluable than the oft-used relative standard deviations of fayalite contents and defines a general metamorphic scale from M0.0 to M1 (where the suffixed number is the rounded *m*). Type 3 chondrites thus span the range M0.0-M0.9 (i.e. 3.0-3.9) and M1 designates type 4. Corresponding applications are then proposed for other chondrite groups (with, e.g., CV secondary classification reduced to essentially three grades from M0.0 to M0.2, that is, subtypes 3.0-3.2). Known type 1 and 2 chondrites are at M0.0 (i.e. the metamorphic grade of type 3.0 chondrites), even so-called "CY" chondrites, since our metamorphic scale is insensitive to brief heating. Independently, we define an aqueous alteration scale from A0.0 to A1.0, where the suffixed number is the (rounded) phyllosilicate fraction (PSF) introduced by Howard et al. (2015). For CM and CR chondrites, the subtypes can be characterized in terms of the thin-section-based criteria of previous schemes (Rubin et al. 2007; Harju et al. 2014) which are thus incorporated in the present framework. The rounding of the PSF to the (in principle) nearest tenth makes the proposed taxonomy somewhat coarser than those schemes, but hereby more robust and more likely to be generalized in future meteorite declarations. We propose the corresponding petrologic subtype to be 3-*PSF*, rounded to the nearest tenth (so that type 1 would correspond to subtypes 2.0 and 2.1). At the level of precision chosen, nonzero alteration and metamorphic degrees remain mutually exclusive, so that a single petrologic subtype ≈ $3+m$-$PSF$ indeed remains a good descriptor of secondary processes (obviating the explicit mention of our separate scales, unless finer subdivisions are adopted for the most primitive chondrites). While proposing a unified way of delineating subtypes, our framework is no substitute for multi-technique studies leading to confidently interpreted multi-parameter schemes for specific groups.

## 1. Introduction

Half a century ago, Van Schmus and Wood (1967) put chondrite taxonomy on its modern foundations by decoupling primary and secondary classifications in a two-dimensional scheme. The primary classification was based on chemical groups presumably representing different primary parent bodies (or families thereof). The secondary classification distinguished petrologic types arranged on a scale from 1 to 6 which characterized parent body processes such as thermal metamorphism, which could vary in degree in the same parent body.

Certainly, the understanding of those processes has evolved. It is no longer believed that the whole ordered sequence from 1 to 6 represents "increasing equilibration" (Van Schmus and Wood 1967), or that chondrules may have formed *in situ* out of type 1 (e.g. CI) material (Mason 1962). Rather, type 1, 2, 3 carbonaceous chondrites—whose labels echoed Wiik (1956)'s types I, II and III—, represent *decreasing* degrees of *aqueous alteration*, and thermal metamorphism increases from type 3 to 6 (e.g. Krot et al. 2014). Although there is thus no *global* ordinal significance in the Van Schmus and Wood (1967) petrologic types, their descriptive value has stood the test of time (Sears 2016) and no meteoriticist can conceivably dispense with them.

With the recognition that type 3 chondrites were the most pristine, it was also quickly realized that they still exhibited considerable diversity (e.g. Dodd et al. 1967; McSween 1977a), and the search for the least processed material warranted further refinement of the classification. Sears et al. (1980) hence subdivided type 3 ordinary chondrites in decimal subtypes ranging from 3.0 to 3.9, a scheme that was later refined by Grossman and Brearley (2005) for the interval 3.00-3.2. Such decimal subtypes (ranging from 3.0 to 3.8) were also proposed for CO carbonaceous chondrites by various authors (Scott and Jones 1990; Sears et al. 1991; Chizmadia et al. 2002; Bonal et al. 2007; Sears 2016). The Meteoritical Bulletin Database (MetBullDB) also lists metamorphic subtypes for Rumuruti chondrites (e.g. Bischoff 2000) and CL chondrites (Metzler et al. 2021). Subtypes have also been proposed for CV chondrites (Guimon et al. 1995; Bonal et al. 2006) and enstatite chondrites (Quirico et al. 2011) but have not gained wide currency.

Type 2 chondrites have also undergone subclassifications. For CM chondrites, Rubin et al. (2007) proposed subtypes from 2.0 (as they renamed type 1) to 2.6, later extended to 2.7 (for Paris; Marrocchi et al. 2014; Rubin 2015) and 2.8-3.0 (Kimura et al. 2020). Harju et al. (2014) proposed subtypes from 2.0 to 2.8 for CR chondrites. Alexander et al. (2013) and Howard et al. (2015) proposed other sets of subtypes, theoretically spanning the range 1.0-3.0, based on the H and phyllosilicate contents, respectively, for these two chondrite groups and some ungrouped carbonaceous chondrites. None of these aqueous alteration schemes has entered the MetBullDB yet (notwithstanding two "CM2.0" (that is, CM1), NWA 11638 and NWA 11732).

This multiplication of subtype definitions developed over the past four decades, sometimes for the same chemical groups, is problematic for the meteoritics student and the classifying community at large. Of course, a classification scheme is a matter of convention and must involve some arbitrariness, but a consensus is difficult to achieve if the delimitation of each subtype obeys no global logic except that of a ranking (as commented by Sears 2016), with divisions sometimes so fine as to make assignment quite uncertain. Scales developed by the same research group for different chondrite groups then stand little chance to be comparable, as Harju et al. (2014) observe when comparing their work on CR with the Rubin et al. (2007) CM classification. For example, a type CM2.5 chondrite according to the scale of Rubin et al. (2007) is more than three times more hydrated than a CR2.5 according to the scale of Harju et al. (2014) (see Fig. 5 of Howard et al. 2015).

Certainly, secondary processing on different parent bodies cannot be wholly described by the same single set of numbers (Sears 2016). Each parent body had its own set of time-temperature histories, water/rock ratios, grain size distribution etc. and different secondary effects will not be correlated the same way, if at all. Thus secondary classification must be ultimately relative (Brearley and Jones 1998; Sears 2016). Yet it would bring welcome clarity if, as for integer Van Schmus and Wood (1967) types, each subtype were defined, even approximately, by *some* simple common empirical property valid across all chondrite groups, warranting a common name (e.g. Howard et al. 2015; Bonal et al. 2006, 2007). Subtypes would then be robust against reclassification to different groups, and could be meaningfully attributed to ungrouped chondrites.

This does not amount to basing secondary classification on one single criterion, especially given the complexity of the secondary processes or perhaps stochastic variability in the protoliths that prevent a simple one-dimensional picture even in single parent bodies. In fact, it is the very correlation between *several* parameters for different members of a group which substantiates their collective interpretation as metamorphic or aqueous alteration effects. A good classification system should take into account different parameters (Sears 2016), thus allowing at times to overrule the indications of single parameters. Although thus the *assignation* of samples to subtypes should rely on several parameters, the *definition* of the subtypes (and the allowed intervals of variation of said parameters) should be ideally tailored, *mutatis mutandis*, after one conventional, well-chosen *defining criterion*.

A difficulty in finding "universal" defining criteria for secondary classification is that, at the level of precision of the modern decimal schemes, thermal metamorphism and aqueous alteration are no longer mutually exclusive (e.g. Schrader et al. 2015; Suttle et al. 2021a; Howard et al. 2010). If, say, a chondrite has been aqueously altered to type 2.5 (according to some scale) and has undergone metamorphism comparable to some type 3.1 chondrite, should the net petrologic type be $2.5 + 0.1 = 2.6$? Obviously not. It thus seems inescapable to define two independent scales for metamorphism and aqueous alteration, with criteria robustly appropriate for each.

The purpose of this paper is thus to propose and justify such a two-dimensional secondary classification for unequilibrated chondrites, with as few arbitrary decisions as possible, building upon the previous literature and the previous classification systems which alone justify the interpretation of secondary effects. Our criteria will be designed to be as simple as possible, so as to ideally appear as natural options to the 21$^{st}$ century meteoritics student not yet familiar with the existing literature on the Van Schmus and Wood (1967) scale. Yet our purpose is *not* to supersede the Van Schmus and Wood scale. In most cases, the essential information *will* still be capturable by one petrological Van Schmus-Wood (sub)type, redefined as a simple function of the metamorphic and alteration degrees, and close to existing practice. So our metamorphic and alteration scales are essentially meant to adjust the definitions of the familiar petrological types. They should not be considered as a complete secondary classification system but rather as a flexible framework which has to be adapted to each group, as we will illustrate for those with existing subclassifications.

## 2. Methods

Although our work is largely based on the literature and mathematical investigations (deferred to the supplements for clarity), it was necessary to measure the olivine compositions of type I and type II porphyritic chondrules in several sections (borrowed from the Muséum national d'Histoire naturelle, Paris) of ordinary and CO chondrites listed in Table 1. The sections were mapped in back-scattered electron (BSE) by a Tescan Clara FEG-dual EDS SEM. Minor and major element concentrations of olivine were obtained with a Cameca SX-5 electron microprobe (EMP) at the Centre de Microanalyse de Paris VI (CAMPARIS). Analytical conditions were 15 kV voltage and 10 nA beam current. Standards were albite (Na), orthose (K, Al), pyrite (Fe, S), $MnTiO_2$ (Mn, Ti), olivine (Si, Mg), $Cr_2O_3$ (Cr), NiO (Ni), apatite (P), vanadinite (Cl). Typical detection limits (see Electronic Annex) were Na: 0.04-0.06 wt%; K: 0.05-0.07 wt%; Fe: 0.1-0.2 wt%; Si: 0.06-0.08 wt%; Ti: 0.04-0.05 wt%; Mg: 0.08-0.1 wt%; Ca: 0.05-0.07 wt%; Mn: 0.08-0.11 wt%; Al: 0.04-0.06 wt%; Cr: 0.04-0.05 wt%; Ni: 0.1-0.2 wt%; P: 0.06-0.07 wt%; Cl: 0.04-0.06 wt%; S: 0.05-0.08 wt%.

We define type I chondrules as those chondrules whose silicates had Mg# = (Mg/(Mg+Fe) <10 mol% upon accretion and type II chondrules as those more oxidized than this. Although diffusion during metamorphism progressively blurs this distinction for the olivine, it is worth remembering that olivine-rich type I and II chondrules can still be readily distinguished texturally (e.g. Scott and Taylor 1983; McCoy et al. 1991), with type I chondrules showing more rounded olivine, pyroxene (still close to pure enstatite because of slower Fe diffusion than in olivine) concentrated near the margin, opaques dominated by metal rather than sulphides, the possible presence of dusty olivine etc. (Fig. 1).

| Sample | | | | Olivine analyses | | | | | | | Point counting (vol%) | | | | | |
| Meteorite | Group | Subtype | Catalog number | Fa$_I$ (mol%) | σ | n | Fa$_{II}$ (mol%) | σ | n | Fa$_I$/Fa$_{II}$ | σ | Type I olivine | Other type I | Type II olivine | Other type II | x$_t$ | σ | n |
| --- | --- | --- | --- | --- | --- | --- | --- | --- | --- | --- | --- | --- | --- | --- | --- | --- | --- | --- |
| Semarkona | LL | 3.0 | 3583sp1 | *1.0* | *0.5* | *15* | *15.7* | *1.9* | *11* | *0.06* | *0.01* | 7.8 | 19.3 | 22.9 | 22.6 | 0.25 | 0.02 | 1000 |
| Krymka | LL | 3.2 | KRYMKAsp | 2.7 | 2.1 | 12 | 21.72 | 8.74 | 11 | 0.13 | 0.03 | 8.3 | 16.2 | 24.2 | 25.8 | 0.26 | 0.02 | 1000 |
| Chainpur | LL | 3.4→3.2 | 1273lm6 | 2.4 | 2.1 | 13 | 19.9 | 5.3 | 17 | 0.12 | 0.03 | 6.2 | 12.8 | 21.9 | 19.4 | 0.22 | 0.02 | 1000 |
| Tieschitz | H/L | 3.6→3.3 | 2772lm1 | 6.4 | 4.1 | 19 | 22.8 | 9.6 | 16 | 0.28 | 0.05 | 6.2 | 15.1 | 20.8 | 26.4 | 0.23 | 0.03 | 1000 |
| Sharps | H | 3.4 | 3575sp1 | 6.7 | 4.1 | 14 | 22.1 | 8.4 | 12 | 0.30 | 0.06 | 9.8 | 16.2 | 21.7 | 21.4 | 0.31 | 0.03 | 1000 |
| Manych | LL | 3.4 | 3128sp1 | 10.0 | 5.9 | 14 | 31.6 | 4.0 | 12 | 0.32 | 0.05 | 11.0 | 20.4 | 23.8 | 25.0 | 0.32 | 0.02 | 1000 |
| Mezö-Madaras | L | 3.7→3.4 | 4416sp1 | 8.9 | 7.0 | 16 | 22.4 | 7.0 | 13 | 0.40 | 0.09 | 5.8 | 12.9 | 22.9 | 31.2 | 0.20 | 0.02 | 1000 |
| Hallingeberg | L | 3.4 | 3776sp1 | 11.5 | 6.0 | 16 | 22.6 | 6.1 | 12 | 0.51 | 0.08 | 9.1 | 22.5 | 21.0 | 20.6 | 0.30 | 0.03 | 1000 |
| Parnallee | LL | 3.7 | 267lm2 | 16.3 | 7.1 | 15 | 25.2 | 5.0 | 10 | 0.65 | 0.08 | 12.2 | 21.8 | 27.6 | 18.6 | 0.31 | 0.03 | 500 |
| Dhajala | H | 3.8 | 3146sp1,2 | 17.6 | 2.9 | 23 | 19.2 | 0.6 | 11 | 0.92 | 0.03 | 5.6 | 12.3 | 25.2 | 23.0 | 0.18 | 0.02 | 1000 |
| Clovis (n°1) | H | 3.6→3.9 | 3352sp1 | 17.6 | 2.4 | 15 | 19.0 | 1.0 | 9 | 0.93 | 0.04 | 6.2 | 17.2 | 19.5 | 16.5 | 0.24 | 0.03 | 1000 |
| Kainsaz | CO | 3.2 | 3745sp1 | 6.9 | 7.3 | 17 | 38.2 | 6.6 | 10 | 0.18 | 0.05 | 13.5 | 28.4 | 4.5 | 2.1 | 0.75 | 0.02 | 2000 |
| Ornans | CO | 3.4→3.2 | 584sp1 | *4.7* | *1.8* | *10* | *34* | *6* | *11* | *0.14* | *0.02* | 14.2 | 18.8 | 5.7 | 1.2 | 0.71 | 0.03 | 1000 |
| Lancé | CO | 3.5→3.2 | 1481sp1 | 6.4 | 6.8 | 18 | 43.0 | 7.9 | 11 | 0.15 | 0.04 | 16.1 | 24.8 | 3.6 | 1.8 | 0.82 | 0.02 | 2000 |
| Warrenton | CO | 3.7 | 743sp1 | 29.1 | 12.7 | 18 | 41.2 | 2.0 | 10 | 0.71 | 0.07 | 13.6 | 18.4 | 6.2 | 1.4 | 0.69 | 0.03 | 1000 |
| Isna | CO | 3.8 | 3239sp1 | 34.2 | 5.2 | 30 | 36.8 | 1.5 | 9 | 0.93 | 0.03 | 10.5 | 19.2 | 5.7 | 1.5 | 0.65 | 0.03 | 1000 |

**Table 1**: Summary of olivine and modal analyses in the samples studied. Arrows in the subtype column indicate suggested reclassifications (the 3.7 classification for Parnallee follows Bonal et al. (2006) and Sears et al. (1991)). For Dhajala, point counting was performed on the (larger) section 3146sp1; the Krymka point counting occurred on (likewise larger) section 3440sp1. The italicized olivine microprobe data of Semarkona and Ornans are from McCoy et al. (1991) and Scott and Jones (1990), respectively.

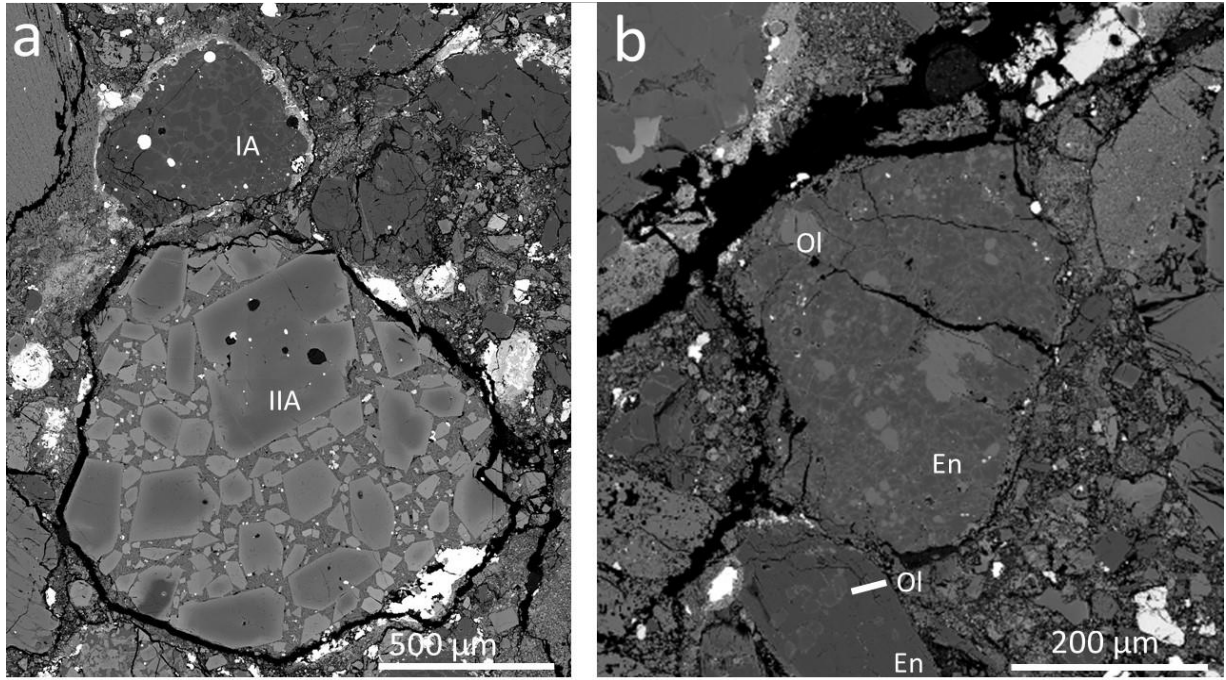

**Figure 1:** Chondrules in the Chainpur (hitherto LL3.4, reclassified LL3.2 herein) meteorite, in back-scattered electron. (a) Prototypical type IA and type IIA chondrules. Little effect of metamorphism is visible. (b) An uncommon type IB chondrule with fayalite-enriched olivine (Fa$_{30-31}$) and more magnesian pyroxene (Fs$_{9-15}$). This chondrule has undergone metamorphic equilibration (analyses are listed in the Electronic Annex).

The difficulty is rather to obtain a *reproducible* average olivine composition in these chondrules, which vary in granulometry, the presence of fractures easing diffusion etc., with of course each individual crystal being zoned and subject to various sectioning geometries. The works of Scott and Jones (1990) and McCoy et al. (1991), on CO and LL chondrites, respectively, which inspired our measurements, did not specify the selection criteria of their chondrules and one may thus suspect a bias in favour of large chondrules with large crystals which diffusion would take longer to equilibrate. To obtain a representative olivine volume-weighted (that is, area-weighted on a representative section) average, a random sampling, uniformly distributed on the surface of the section, would be the cleanest way to go (with labelling each point as either type I or type II) in principle. For instance, Dodd et al. (1967) selected points in 500 µm steps in traverses 1000 µm apart. So one could envision applying this method and identify the chondrule type analysed. Yet, even allowing for a 20 µm leeway to find neighbouring silicate grains (Dodd et al. 1967), such a method would be inefficient to find olivine (rather than pyroxene or even other phases), and *a fortiori* olivine in sufficient number for each chondrule type (e.g. if one of them is relatively rare). For example, in ordinary chondrites, type I olivine (most critical for classification) would be reached in less than a tenth of the time, according to our point counting estimates (discussed in supplement E).

So we chose instead to generalize, *mutatis mutandis*, the method introduced by Grossman and Brearley (2005) to estimate the first moments of the $Cr_2O_3$ contents in type II chondrule olivine, that is, analyze the core of one "typical olivine" per randomly chosen chondrules. Supplement A.1 shows the statistical advantage of performing only one analysis per chondrule for a given total number of analyses.

We performed the "random selection" by making 1000 µm steps (the exact step being immaterial so long it is predefined and larger than typical chondrules) and choosing the closest suitable chondrule of the quota of interest (porphyritic type I, and then type II). As a crude way to introduce weighing by olivine amount, we imposed that olivine should dominate over pyroxene. This is necessary because pyroxene-rich type I (IBA, IB) chondrules may originally contain distinctly more ferroan olivine than type IA chondrules (e.g. Jones 1994; Jacquet et al. 2021) and the smaller size of the olivine chadacrysts would further facilitate the diffusion of Fe during metamorphism. So we avoided additional variability from a likely minor contributor in terms of total olivine volume. Also, the pyroxene-rich type I and type II chondrules may also be less easily texturally distinguishable in ordinary chondrites.

To make the concept of "typical olivine" more objective, we defined it by mentally reordering all apparent (unbroken) olivine crystals of the chondrules in order of decreasing apparent area and taking the grain for which the cumulative olivine area reaches half the total area. If this grain contained an apparent relict (e.g. a forsteritic grain inside a type II chondrule or a dusty olivine in a type I chondrule), another one closest in size was chosen. If the typical grain is the largest olivine grain of the chondrule (that is, if that olivine makes more than 50 % of the total apparent area of the chondrule olivine), the whole chondrule was deemed unsuitable and the next nearest one was sought. This excludes isolated olivine grains (Jacquet et al. 2021) or any unrepresentative chondrule fragment where the typical grain as defined above cannot be trusted to be typical of the unbroken chondrule. This removes any ambiguity of definition with respect to the matrix and mitigates the effect of fragmentation (either post- or pre-accretion), unless it affected preferentially one subpopulation of type I or type II chondrules[1]. Of course, the identification of the typical olivine during a microprobe session is, in practice, subject to error (for we obviously do not actually measure the surface of all olivine grains on the fly), but keeping our definition in mind does reduce the freedom of choice (and therefore the resulting contribution to the standard deviation, see Supplement A).

Strictly speaking, the average olivine composition of a chondrule type calculated this way is *not* the olivine-volume-weighted average composition of olivine in this chondrule type. Although the "typical grain" may yield a fairly accurate estimate of the average olivine composition of a given chondrule (as discussed mathematically in Supplement A.2), the chances of a given chondrule to be chosen, which are proportional to the area of its Voronoi domain (i.e. the set of points where it is the closest suitable chondrule), are not exactly proportional to its amount of olivine. Yet barring a strong correlation between typical Fa content and chondrule size, our calculated average should not be far from the olivine volume-weighted one, and as such, the method is quite simple to implement and merely regulates the already standard initiative of analysing chondrule silicates during the classification of a chondrite. By generalizing the Grossman and Brearley (2005) method, it allows all metamorphic subtypes of type 3 chondrites to be determined with the same kind of dataset, with only the required statistics of each chondrule type varying with the suspected level—clearly, we need more type II chondrule data for lower than for higher subtypes, where their olivine is nearly equilibrated. We suggest at least 20-30 type I olivine analyses for putative type 3.2-4 samples. In this study, we measured 23-39 data points per section.

### 3. Overview

In the same way Stöffler et al. (1991) called their *S*hock stages "S1", "S2", etc. and Wlotzka (1993) his *W*eathering grades "W0", "W1", etc., we will introduce a prefix "A" for *A*queous alteration grades and a prefix "M" for *M*etamorphic grades. The prefixes at once prevent confusion with any other scale in the literature.

---

[1] Nevertheless, type II chondrules may be so rare in carbonaceous chondrites that IOG must be included for meaningful statistics for types 3.0-3.2 determination (Grossman and Brearley 2005).

The aqueous alteration degrees will run from A0.0 to A1.0, with A0.0 corresponding to type 3 chondrites, and A0.9 and A1.0 to petrologic type 1. Type 2 chondrites encompass the intermediate alteration grades from A0.1 to A0.8. For type 1 and 2 samples, the number after the prefix is intended as the number to be subtracted from 3 to obtain the Van Schmus-Wood petrologic subtype.

The metamorphic grades will run from M0.0 to M1, with M1 standing for petrologic type 4 (and higher). For type 3 samples, the number after the prefix M is intended as the decimal part of a redefined petrologic type (e.g. M0.2 should correspond to 3.2). In the discussion to follow though, petrologic types will refer to past literature, unless otherwise noted, again to prevent confusion.

The purpose of the succeeding sections is to provide general definitions for the aqueous alteration and metamorphic grades. More precisely, we wish to define real numbers *a* and *m* (which we may call aqueous alteration and metamorphic *degrees*) whose rounding to a neighbouring tenth will yield the numbers after the prefixes A and M, respectively (i.e. we have A$r(a)$ and M$r(m)$ with $r$ the rounding function). We will find that $a = PSF$ (phyllosilicate fraction) and $m = Fa_I/Fa_{II}$ (ratio of average Fa contents of type I and II chondrules) are generally good options. By default, the rounding should be to the *nearest* tenth (e.g. if $m = 0.44$, we shall have M0.4), but it can be adapted for each group (e.g. to mark petrographically significant transitions). For a particular sample, the rounding may be further amended if other assignation criteria overrule it. Again, choosing a *defining criterion* does not mean that *assignation criteria* (i.e. those used, in practice, to assign a specimen to a grade) should be restricted to it (Sears 2016). Kimura et al. (2024), for example, proposed 20 criteria to delineate the 3.0 subtype, which will largely fit our own M0.0/M0.1 and A0.0/A0.1 boundaries. The defining criterion, as we put it, merely constrains the boundaries of the different assignation criteria for the defined grades, and the more diverse such assignation criteria are, the more robust the secondary classification is.

## 4. The aqueous alteration scale

We start with the aqueous alteration scale, because it is on the whole most straightforward. Indeed Howard et al. (2015) made a convincing case that the phyllosilicate fraction (PSF) allowed cross-calibration of aqueous alteration scales for different chondrite groups. They did not include other putative secondary minerals (like sulfides, oxides, carbonates, fayalite…) which may be controversial; their modal contribution would be anyway comparatively minor. The PSF is the (volume) ratio of (hydrous) phyllosilicates to all silicates. To be sure, the PSF is affected by the chondrule/matrix ratio, since the matrix is more easily hydrated (Abreu 2016), but so long the alteration grades are sufficiently wide to overcome effects of such fluctuations within a given chemical group, the PSF should do as a relative scale within it.

We thus adopt $a = PSF$ as the defining criterion for our alteration grades. Russell et al. (2021)'s definition of type 1 carbonaceous chondrites by PSF > 0.9 is conformable to our convention of type 1 corresponding to A0.9 and A1.0 (although our convention may be somewhat wider if PSF ≥ 0.85 warrants a rounding to A0.9). Conversely, the restriction of type 3 to A0.0 keeps all CV, CO, UOC at type 3 and most CRs at type 2. The type 2/type 3 transition would roughly correspond to the detectability of phyllosilicates by X-ray diffraction and infrared spectroscopy (5 and 2 vol% respectively; Krämer-Ruggiu et al. 2022). It may be noted that the Van Schmus and Wood (1967) definitions of types 1 and 2 were appropriate for CI and CM chondrites, respectively, and are now obsolete given the increasing diversity of aqueously altered chondrites known.

While the PSF is thus the defining criterion for the aqueous alteration degree, it certainly does not provide a full classification system alone. Howard et al. (2015) acknowledged that the Position Sensitive Detector X-Ray Diffraction (PSD-XRD) which allows to measure it (and whose data will be used here, unless otherwise noted) is not routine. Other methods exist (conventional XRD, which is less sensitive, possibly thermal gravimetric analysis combined with infrared spectroscopy; Krämer-Ruggiu et al. 2022) which are not either. Thus more operational, thin section-applicable assignation criteria must be sought, and in fact have been offered in the literature. These we now undertake to integrate for different chemical groups. We note that the general Kimura et al. (2024) criteria for the 3.0/2.9 boundary, on alteration of chondrule feldspar and glass, CAI melilite and matrix amorphous phases should fit our A0.0/A0.1 boundary, although only that on the destruction of O-anomalous presolar grain (dropping below 100 ppm) is quantitative.

### 4.1 CM chondrites

Aqueous alteration has been best studied in CM chondrites. Rubin et al. (2007) and Rubin (2015) devised a petrographic classification scheme based on chondrule silicate alteration, metal abundance, matrix and tochilinite-cronstedtite intergrowths (TCI) compositions, and the nature of sulfides and carbonates. This classification was found to correlate well with the PSF (Howard et al. 2015; Fig. 2). So, essentially, the matter for us is simply to redraw the boundaries with the same diagnostic criteria.

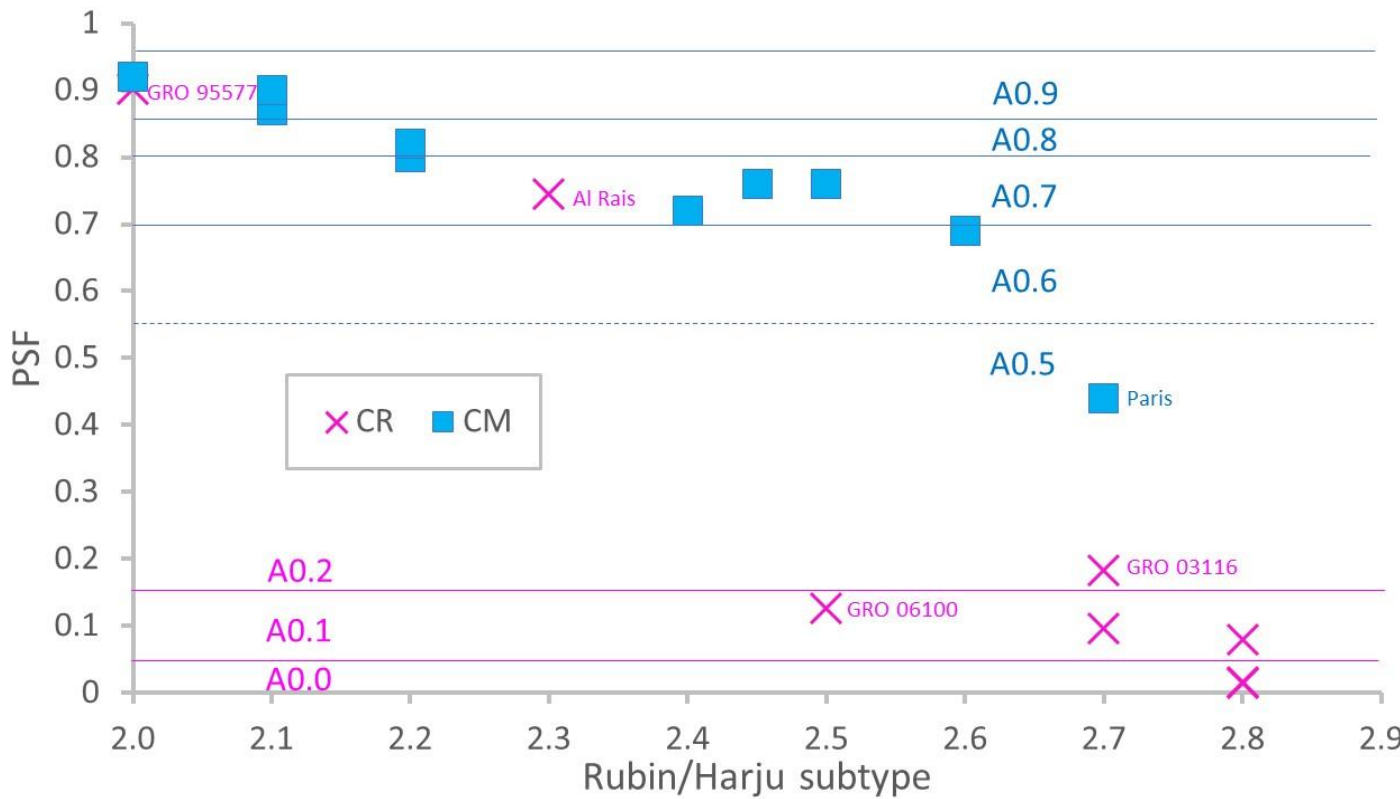

**Figure 2**: PSF (Howard et al. 2015) as a function of the subtypes proposed by Rubin et al. (2007) and Harju et al. (2014) for CM and CR chondrites, respectively. We also show the boundaries chosen for the alteration grades A0.5-A0.9 (only tentative for A0.5/A0.6) for CM and A0.0-A0.2 for CR. Very altered (e.g. Marrocchi et al. 2023) or shock-heated (Briani et al. 2013) CRs are explicitly named.

Our freedom is limited to adjustments of the rounding of the PSF. We tie it further by making our alteration grades conservative with respect to the Rubin et al. classification, in that each of them corresponds to one or two Rubin (2015) subtypes, so as to make conversion straightforward. We note that PSF ~ 0.8 corresponds to an important transition, namely the beginning of significant replacement of mafic phenocrysts in chondrules, first affecting mostly pyroxene (most rapidly replaced; e.g. Hanowski and Brearley (2001); see also Fig. 7 of Howard et al. (2015)). This transition means that the value PSF=0.8 should be a boundary rather than the center of an alteration grade interval. Thus we elect to have A0.8 correspond to PSF between 0.8 and 0.85 (rather than the interval [0.75; 0.85[), encompassing Rubin et al. (2007)'s subtypes 2.3 and 2.2, whereas A0.7 corresponds to their 2.4 and 2.5.

Table 2 lays out all our alteration grades adjusted for CM and the relationship with the Rubin classification is illustrated in Fig. 2. This effects a synthesis between the Rubin et al. (2007) and Howard et al. (2015) classifications. We have added criteria of Velbel et al. (2015) on chondrule enstatite and ferroan olivine, and adjusted the percentage limits for the chondrule mafic phenocrysts replacement for A0.8 (10 and 50 % rather than 2 and 85 %) to make them more readily evaluable and in line with the suggested PSF limits (this does not change the classification of the specific chondrites studied by Rubin et al. (2007))[2]. Our table does not include the weakly altered "CM2.8", "CM2.9" and "CM3.0" chondrites of Kimura et al. (2020) since their PSF is unknown and their affiliation to CM may be uncertain, given the large difference in alteration degree. It is also premature to list classification criteria for subtypes populated by only one member; in fact such is already the case for our A0.5 (Paris) whose properties are here only tentatively taken from Rubin (2015)'s subtype 2.7. At the other extreme, the grade A1.0 is unrepresented in CMs since their highest PSF measured by Howard et al. (2015) was 0.93[3]. CM1 would be restricted to grade A0.9 chondrites (which the MetBullDB variously refers to as "CM1", "CM1/2" or even "CM2" currently).

---

[2] We have also « interpreted » the vague "~1 vol%" metal of their subtype 2.6 (our A0.6) as "0.3-1.5 vol%".
[3] Thus Rubin et al. (2007)'s subtype 2.0 is likely not actually populated if its requirement of less than 1 % of surviving chondrule phenocrysts is taken literally. Such a threshold must indeed be difficult to evaluate in optical or electronic microscopy.

| Alteration grade | A0.5 | A0.6 | A0.7 | A0.8 | A0.9 |
|---|---|---|---|---|---|
| Corresponding classification | CM2.5 | CM2.4 | CM2.3 | CM2.2 | CM2.1 (=CM1) |
| Example | Paris | QUE 97990 | Murchison | Nogoya | MET 01070 |
| Rubin (2015) subtype | 2.7 | 2.6 | 2.5-2.4 | 2.3-2.2 | 2.1-2.0 |
| Howard et al. (2015) subtype | (2.0) | 1.9-1.6 | 1.6-1.4 | 1.4-1.3 | 1.3-1.1 |
| Velbel et al. (2015) stage | 0 | 0 | 1-2 | 3-5 | 6 |
| McSween (1979) | N/A | N/A | partially altered | altered, highly altered | N/A |
| PSF | (0.44) | 0.55-0.7 | 0.7-0.8 | 0.8-0.85 | 0.85-0.95 |
| Matrix silicates | phyllosilicates + amorphous | phyllosilicates | phyllosilicates | phyllosilicates | phyllosilicates |
| matrix MgO/"FeO" | 0.35-0.43 | 0.35-0.43 | 0.35-0.43 | 0.5-0.7 | 0.5-0.7 |
| matrix S/SiO$_2$ | 0.1-0.18 | 0.1-0.18 | 0.1-0.16 | 0.07-0.08 | 0.05-0.07 |
| Large TCI clumps (vol%) | 5-20 | 15-40 | 15-40 | 15-40 | 2-5 |
| TCI "FeO"/SiO$_2$ | 4.0-7.0 | 2.0-3.3 | 1.5-3.3 | 1-2 | 1-1.7 |
| TCI S/SiO$_2$ | 0.40-0.60 | 0.18-0.35 | 0.14-0.35 | 0.05-0.2 | 0.05-0.09 |
| Carbonates | Ca carbonates | Ca carbonates | Ca carbonates | Ca carbonates | complex and Ca carbonates |
| Dominant sulfides | po+pn | po+pn | po+pn | po+pn+int | pn+int |
| Chondrule mesostases | phyllosilicates | phyllosilicates | phyllosilicates | phyllosilicates | phyllosilicates |
| Chondrule mafic silicate alteration (%) | 0 | 0 | <10 | 10-50 | >50 |
| Chondrule enstatite/ferroan olivine | unaltered | unaltered | incipient alteration | advanced replacement | mostly replaced |
| Metal Fe-Ni (vol%) | >1 | 0.3-1.5 | 0.03-0.3 | 0.03-0.3 | <0.03 |

**Table 2**: Aqueous alteration scale for CM chondrites. Parenthesized PSF value under A0.5 is that of its sole member, Paris (see Howard and Zanda 2019), making any delineation of the grade premature. po = pyrrhotite, pn = pentlandite, int = intermediate sulfide. Most petrographic criteria from Rubin (2015), with MgO/"FeO" from Kimura et al. (2020) and replacement of enstatite and ferroan olivine after Velbel et al. (2015).

We commented above that our classification is coarser than Rubin et al.'s but it is hereby more robust and practical. Most CM chondrites are genomict breccias, prone to significant variations in apparent alteration degree distributions from one thin section to another (e.g. Lentfort et al. 2021). So intrinsic variations in PSF are superimposed on analytical errors (5 vol%; Howard et al. 2015). It is thus pointless to base a classification on PSF variations below 0.05. Now, variations beyond this scale *are* meaningful; there is no way, for example, that a A0.9 (i.e. with chondrule phenocrysts mostly replaced) can be mistaken for a A0.7 (where these phenocrysts are largely intact, and TCI are BSE-brighter; Lentfort et al. 2021), or vice-versa, unless the studied thin section mainly samples one unrepresentative clast. Thus in our scheme, uncertainty cannot be more than one subtype (for borderline samples). We note that the PSF of a genomict breccia still makes sense in representing the silicate-weighted average of the PSF of its constituent lithologies; one could thus display both the bulk and range of alteration degrees (for lithologies representing at least 10 vol%; Jacquet 2022) of a CM chondrite, e.g. in the form "A0.7 (0.7-0.8)". We note that PSD-XRD measurements such as those of Howard et al. (2015) (on 0.2 g aliquots of homogenized >>1 g samples) "only" bear on the whole-rock PSF, but Rubin et al. (2007)'s thin section criteria embodied in Table 2 would allow estimates at the lithology scale (and thus of the range of subtypes represented).

Besides, our taxonomic simplification is only for "first-look" classification purposes (upon declaration to NomCom). Nothing here prevents any researcher interested in subtler insights on alteration from plotting any property against any underlying alteration metric (e.g. TCI composition, PSF…) taken as a continuous variable, rather than the discrete subtypes.

### 4.2 CR chondrites

The CR chondrites were subclassified based on petrographic properties by Harju et al. (2014). The PSF sets apart the same grouplet of heavily aqueously altered CRs (Al Rais and GRO 95577) which they ranked at their lowest subtypes (2.3 and 2.0, respectively). Weisberg and Huber (2007) did not report anhydrous silicates in GRO 95577, although Howard et al. (2015) measured PSF = 0.89 (sufficient for a CR1 classification though), but remnants of olivine phenocrysts unreplaced by phyllosilicates might be discernable in their Fig. 2b. It would be premature to delineate the one-member alteration grades corresponding to these meteorites, not to mention the uncertainty on their genetic link with mainstream CRs (Marrocchi et al. 2023).

As to the other, "mainstream" CRs, the Harju et al. (2014) sequence is not always consistent with water content-based classification (Alexander et al. 2013) or TEM examination of matrices (Abreu 2016). Many of the Harju et al. (2014) subtypes differ only in subjectively defined degrees (e.g. mafic phenocryst alteration may be "none", "incipient", "moderate", "altered"), which may be also liable to sample heterogeneities (Abreu 2016). In fact, from a PSF perspective, there is not much variation after all (as held from a textural perspective by Le Guillou et al. 2015). Indeed, when excluding Al Rais and GRO 95577, the measured PSF do not exceed 0.25 (Dhofar 1432; Howard et al. 2015). So there is only room for three alteration grades for mainstream CRs, viz. A0.0, A0.1 and A0.2.

Alteration grade A0.0 would be carved out of Harju et al. (2014)'s subtype 2.8. After deducting amorphous phases, Abreu and Brearley (2010)'s CR3 MET 00426 and QUE 99177 show PSF=0.01 and 0.02, respectively. Certainly, these meteorites were not exempt from all aqueous alteration (e.g. the "smooth rims" after silica-rich igneous rims; Harju et al. 2014; Martinez and Brearley 2022), but no known carbonaceous chondrite is (even Acfer 094, with its "cosmic symplectites"; e.g. Sakamoto et al. 2007), and their presolar grain abundances bear witness to their pristinity (e.g. Haenecour et al. 2018; Leitner et al. 2015), using Kimura et al. (2024)'s O-anomalous grain criterion. While Harju et al. (2014) consider the mostly crystalline state of chondrule mesostases as an alteration effect, it may be primary, e.g. as a result of slow cooling (e.g. Jacquet et al. 2021), similar to that proposed by Wick and Jones (2012) for the plagioclase-bearing chondrules in CO chondrites. We thus concur with their literature classification as CR3 (Abreu and Brearley 2018).

It may be noted that Howard et al. (2015) eventually lumped amorphous silicates with phyllosilicates in the calculation of the PSF. Since the latter originally grow at the expenses of the former, this would hide the relative pristinity of the CR3 above (all CR would then have PSF > 0.1; see Fig. 4 of Howard et al. 2015), as also observed by Abreu (2016). We thus hold that, true to its etymology, the PSF numerator should not extend to amorphous phases, regardless of whether they were hydrated prior to accretion or at the very onset of parent body alteration (e.g. Le Guillou et al. 2015).

Most CRs would be A0.1, which encompasses a wide range of Harju et al. (2014) subtypes. A0.2 would essentially correspond to the alteration degree of Renazzo, i.e. Harju et al. (2014)'s subtype 2.4, with their subtype CR2.5 GRA 06100 having undergone brief reheating (Briani et al. 2013) and thence partial dehydration (Prestgard et al. 2023). There is, however, no PSF datum for meteorites classified at Harju et al. (2014)'s subtype 2.4. Yet phyllosilicates in those studied by Abreu (2016) account for somewhat over half of their matrix, which in turn have modal abundances 25.8-37.6 vol% so, with only moderate chondrule alteration, this should yield a PSF around 0.2. Direct measurements would be worthwhile.

Table 3 lists our proposed alteration scale for CRs. It, again, can be seen as a synthesis between the Harju et al. (2014) and the Howard et al. (2015) scales (see Fig. 2), though not as conservative as our CM classification as it divides Harju et al. (2014)'s subtype 2.8. The grades A0.0, A0.1, A0.2 also correspond to Prestgard et al. (2023)'s degrees (i), (ii) and (iii), respectively, and would respect (if simplify) the ranking of Abreu (2016), notwithstanding the contradictions on (possibly heterogeneous) LAP 04516 in the literature, with Harju et al. (2014) and Prestgard et al. (2023) having seen less hydrated samples than Abreu (2016).

| Alteration grade | A0.0 | A0.1 | A0.2 | A0.7 | A0.9 |
|---|---|---|---|---|---|
| Corresponding classification | CR3 | CR2.9 | CR2.8 | CR2.3-an | CR1-an |
| Examples | MET 00426, QUE 99177 | Acfer 187 | Renazzo | Al Rais | GRO 95577 |
| Harju et al. (2014) subtype | 2.8 | 2.8-2.6 | 2.5-2.4 | 2.3 | 2.0 |
| PSF | 0-0.05 | 0.05-0.15 | 0.15-0.25 | (0.74) | (0.90) |
| Matrix silicates | mostly amorphous | mostly amorphous | mostly phyllosilicates | mostly phyllosilicates | mostly phyllosilicates |
| O-anomalous presolar grains in matrix | >100 ppm | 50-100 ppm | <50 ppm | <50 ppm | <50 ppm |
| Chondrule mafic phenocryst alteration | none | none | moderate | altered | altered |
| Magnetite visible? | no | no | yes | yes | yes |

**Table 3**: Aqueous alteration scale for CR chondrites. Chondrule criteria from Harju et al. (2014); matrix silicate mineralogy after Abreu (2016); PSF after Howard et al. (2015); presolar grain abundances after Haenecour et al. (2018); Leitner et al. (2015), adopting the Kimura et al. (2024) 3.0/2.9 threshold as our A0.0/A0.1 boundary. Parentheses indicate that the alteration grades in question are only represented by one member so cannot be meaningfully delineated.

### 4.3 Other carbonaceous chondrites

Phyllosilicates are detected in a few CO chondrites (Alexander et al. 2018), with the highest PSF value of 0.05 for CO3.0 ALH A77307. We thus deem them all A0.0.

Among CV chondrites, phyllosilicates are reported in the so-called oxidized Bali-type subgroup ($CV_{oxB}$; as opposed to the oxidized Allende-type $CV_{oxA}$; e.g. Weisberg et al. 1997). With a PSF = 0.04 (Howard et al. 2010), this warrants again a type A0.0. One might consider subdividing A0.0 in a A0.00 and A0.05 (for PSF lower and higher than 2.5 %, respectively), with the latter representing $CV_{oxB}$ which thus would no longer need a specific designation among $CV_{ox}$ (see also Gattacceca et al. 2020).

All CI chondrites studied by King et al. (2015) are A1.0, as anhydrous silicates, likely chondrule and AOA debris (e.g. Morin et al. 2022), are very minor.

The so-called "CY chondrites" (e.g. Ikeda 1992; King et al. 2019, 2021; Suttle et al. 2021a, 2023) pose a problem of convention. Indeed, they presently contain no detectable hydrous phyllosilicate but this may be the result of a brief

(possibly impact-related) heating, judging from Raman properties and olivine zoning (Suttle et al. 2023; Nakato et al. 2008). Their previous PSF may have reached from 0.6 to 1.0 (i.e. CM- to CI-like; Suttle et al. 2021a). If we sticked to the *present-day* PSF, all CY should be A0.0, that is type 3, but this would falsely convey a pristine state to most meteoriticists. Suttle et al. (2021a) rather attributed petrologic types according to the inferred pre-heating state, distinguishing "CY1" and "CY2" chondrites. This corresponds to the usage in the MetBullDB where "CY1" are listed as "CI1" or "CI1/2" and "CY2" as "C2-ung". So the relevant PSF could be the *maximum* one inferable in preterrestrial alteration history. This has the drawback of being more model-dependent, but the inferred thermal reheating may be explicitly indicated, e.g. in the form of the suffix "T" proposed by Tonui et al. (2014), e.g. "A1.0T" or "CY1T"[4] (which we may supplement with a dedicated thermal classification, to which we will return in the section The metamorphic scale). We comment that the "CY1T" and "CY2T" may not be genetically related (to the same preaccretionnary protolith) as the former show distinctly higher "plateau" volatile element abundances, in line with their higher matrix proportion (see Fig. 13 of King et al. (2019)[5]). So the chondrule-free CY1T probably never contained many chondrules (similar to CI; Morin et al. 2022), and may form in due course a distinct (primary) group, if not related to CI themselves.

4.4 Noncarbonaceous chondrites

Aqueous alteration effects, and in particular phyllosilicates, are detected in unequilibrated ordinary chondrites (e.g. Hutchison et al. 1987; Brearley 2014). For Krymka and Bishunpur, Grant et al. (2023) found 12 and 16 wt%, respectively, of an "X-ray amorphous" component, which they tentatively identified as phyllosilicates, but these cannot make the whole of it as (i) their specific diffractions peaks were missing (ii) chondrule mesostasis must also be present, at modal abundances presumably comparable to the plagioclase modes of equilibrated ordinary chondrites (8.9-9.8 wt%; Dunn et al. 2010) (iii) their estimated matrix-normalized water abundance (4.8-8.2 wt%) is far below their values for pure phyllosilicates (15-20 wt%) which must thus make a corresponding fraction only of the matrix (itself 10-15 wt% of OCs). Thus bulk phyllosilicate abundances must be a few wt% for the least metamorphosed ordinary chondrites, similar to CVs, and all OCs should thus be A0.0.

---

[4] Prestgard et al. (2023) inserted an hyphen, as in "CR2-T" but we reserve the hyphen to lithology mixing and "an" and "ung" suffixes (Jacquet 2022). We note that "T" has been proposed as an abbreviation for tafassites (Ma et al. 2022). It is thus best to tie the "T" suffix immediately to the petrologic type.

[5] The dehydration event must have incurred an additional depletion of elements more volatile than Te (e.g. Cd; Nakamura 2005).

Same for most R chondrites. Yet R5 chondrite LAP 04840 contains 13.4 vol% ferri-magnesiohornblende and 0.4 vol% phlogopite, which may result from high-temperature hydration (McCanta et al. 2008). Similarly, R6 chondrite MIL 11207 contains 13 vol% OH-bearing minerals (Gross et al. 2017). Although amphiboles are not phyllosilicates, and notwithstanding our focus on type 1-3 chondrites, we could extend the numerator of the PSF (perhaps to be renamed "Nominally Hydrous Silicate Fraction") to such nominally hydrous silicates. In this case, those equilibrated chondrites would thus warrant a A0.1 grade.

## 5. The metamorphic scale

Since the Sears et al. (1980) scheme for ordinary chondrites is widely used, with, in fact, no recent rival to our knowledge (since the Grossman and Brearley (2005) refinements were conservative), any redefinition of the metamorphic degrees would do well to closely stick to their subdivisions for ordinary chondrites. Yet, what defining criterion should we abstract from the OC subtypes? Sears et al. (1980) originally calibrated their subclassification (and the range of variation of their assignation criteria) after the increase of the induced thermoluminescence (TL) sensitivity, as a result of the crystallization of secondary feldspar out of chondrule mesostases. Specifically, their petrologic types PT may be given by:

$$PT = 3.8 + 0.1 \times \text{floor}(3 \log_{10} TL_{\text{Dhajala}})$$

With "floor" the floor function[6] and $TL_{\text{Dhajala}}$ the induced TL sensitivity at the ~130 °C peak, normalized to the Dhajala H3.8 chondrite.

However, the same formula applied to CO chondrites would make ALH 77307 (CO3.0) a type 3.3 (Scott and Jones 1989). This may be due to the presence of (feldspar-bearing) refractory inclusions or chondrules rich in primary anorthite (Wick and Jones 2012), so Sears et al. (1991a) in effect replaced the factor 3 of the above formula with a factor 4 for CO chondrites. Conversely, CK chondrites, which are certainly metamorphosed, show no detectable induced TL, for as yet unclarified reasons (Guimon et al. 1995). For enstatite chondrites, TL sensitivity does not bear a simple relationship to metamorphism either (Zhang et al. 1996). Clearly, TL is not solely a function of thermal metamorphism (when it is) and although powerful to resolve subtypes within given chemical groups, including COs (Sears et al. 1980; Sears 2016), it was not claimed by its developers to provide the universal scale we are seeking (e.g. Sears 2016).

---
[6] Which returns the highest integer smaller than its argument (e.g. floor(2.4)=2).

Bonal et al. (2006, 2007) sought a universal sensor of the peak metamorphic temperature in the maturity of organic matter (previously studied in UOCs by Quirico et al. 2003) but, however relevant as a relative scale, this was criticized by Sears (2016), in particular as to the uncertain premise that the precursor organic matter was the same across all chondrite groups. At times, because of its sensitivity to peak temperatures, it may measure rather shock events than the protracted metamorphism of interest here (e.g. Briani et al. 2013). The Bonal et al. (2007) CO subtypes would also jump from 3.1 to 3.6, hinting at an overprecise classification in between, only evaluable with Raman. Also, no more than TL sensitivity does Raman spectroscopy lend itself to a "natural" expression of $m$ without arbitrary numerical parameters. Finally, beyond subtype ~3.6, organic matter maturity may provide little discrimination (see e.g. Fig. 2 of Bonal et al. 2016). This, again, does no prejudice to its important classifying power for less metamorphosed chondrites (it may be the handiest tool for some groups such as CVs, as discussed in their own subsection later).

Olivine phenocrysts in chondrules have a better claim as "universal" sensor of metamorphism, in addition to being, in practice, the criterion most used by OC and CO classifiers. As noted by Grossman and Brearley (2005), olivine is the phase most resistant to aqueous alteration, and so long overgrowths possibly precipitated from fluids (e.g. Han et al. 2022) or serpentinized areas are avoided, its properties are independent of aqueous alteration. Their mineral chemistry should be likewise independent of shock effects (again for whatever portion of olivine escaped phase transitions such as melting) given the short diffusion lengths then (e.g. a few microns in shocked L chondrites; Ciocco et al. 2022).

In fact, the very type 3/type 4 transition *was* defined by Van Schmus and Wood (1967) in terms of the homogenization (at the 5 percent mean deviation (PMD)[7] of FeO level) of olivine—which olivine, in ordinary chondrites, mainly meant chondrule olivine, because of the low abundance of matrix and the technical difficulty in analyzing small grains (Grossman 2011). The *progress* toward this boundary could thus as well be measured by the degree of homogenization of olivine. In their table 4, Sears et al. (1982) showed that their subtypes could be correlated with the fayalite relative standard deviation *RSD(Fa)*, i.e. the standard deviation of the fayalite content divided by the mean (often called the "coefficient of variation"), which we can subsume in the following equation:

$$PT = 3.9 - 0.1 \times \text{floor}(10\,\text{RSD(Fa)})$$

---

[7] The PMD is the average relative absolute deviation to the mean, expressed as a percentage. By virtue of the Cauchy-Schwarz inequality, it is smaller than the relative standard deviation.

If generalized, this would suggest a simple $m \approx 1 - RSD(Fa)$. This, in fact, is the formula commonly used for R chondrites (Bischoff 2000). Yet this does not quite do, for (i) this formula actually overestimates the petrological types assigned to ordinary chondrites in the literature below subtype 3.6 (e.g. Fig 3; Table 5 of Sears et al. 1982; Table 4 of Sears et al. 1991b), warranting revision, and (ii) the *RSD(Fa)* actually stalls around 50 % for OCs below subtype 3.5 (e.g. Sears et al. 1982). At this stage, the relative standard deviation is essentially indistinguishable from that of the original assemblage, which ultimately depends on the type I/type II chondrule ratio (see Supplement C). In CO chondrites, *RSD(Fa)* reaches 125 % for the least metamorphosed sample ALH 77307 (Scott and Jones 1990). Also *RSD(Fa)* depends on what olivine grain size can be selected (e.g. in the matrix) and the spatial resolution of analyses; it is as such most subject to inter-operator bias. Wherever type II chondrules are rare, as in CV chondrites, the PMD (e.g. McSween 1977b) and *a fortiori* the *RSD(Fa)* may be strongly affected by any fluctuation in the number of type II olivine analyses selected.

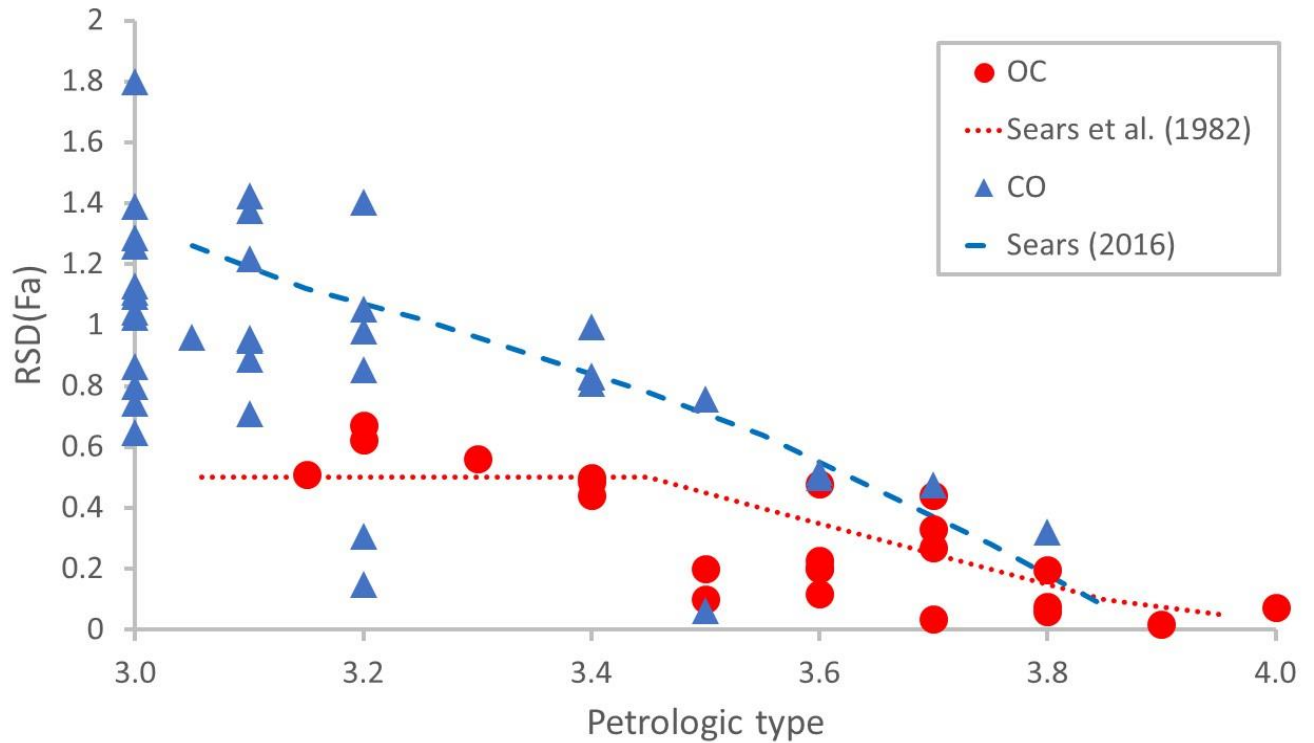

**Figure 3**: RSD(Fa) as a function of petrologic type (data from Scott and Jones (1990) and MetBullDB for CO; Sears and Hasan (1987), Scott (1984), Dodd et al. (1967) for UOC). We also show the RSD(Fa) values proposed by Sears et al. (1982) for UOC, and Sears (2016) for CO.

Now, the stability of the relative standard deviation below 3.5 for ordinary chondrites does not mean that the fayalite distribution does not change before. The Fe enrichment of type I chondrule olivine is in fact already appreciable from subtype 3.2 onward (e.g. McCoy et al. 1991; Huss et al. 2006). In fact, with Fa contents of type II chondrules showing smaller changes during metamorphism (e.g. McCoy et al. 1991; Scott and Jones 1990), the main control on *RSD(Fa)* is type I chondrule composition. Thus type I chondrule olivine composition should be essentially the metric of metamorphism. It was incidentally the basis for the calibration of the Scott and Jones (1990) scheme for CO chondrites (matching Warrenton and Parnallee as their then-current subtype 3.6).

The simplest function of the mean[8] fayalite content of type I chondrule olivine (henceforth $Fa_I$) going from 0 to 1 as metamorphism advances to type 4 would be a linear function (for a given chemical group), ideally:

$$m = \frac{Fa_I - Fa_{I,0}}{Fa_{limit} - Fa_{I,0}}$$

With $Fa_{I,0}$ the value before metamorphism, and $Fa_{limit}$ the value of an hypothetical analog carried past olivine equilibration (i.e. type 4).

This will be our guiding idea. However, the above expression is not yet an operational formula, for $Fa_{I,0}$ and $Fa_{limit}$ would have to be inferred for any single chondrite, which is somewhat model-dependent (e.g. as to taxonomic assignations), especially for an ungrouped chondrite. $Fa_{limit}$ may not be well-defined if no part of the parent body reached type 4, and even if it did, there may be a finite range of values therein (e.g. Rubin 1990).

Yet, whenever type II chondrules exist in significant amounts (that is all but the most reduced chondrites), $Fa_{I,0} \ll Fa_{limit} \sim Fa_{II}$ with $Fa_{II}$ the mean type II chondrule composition. Indeed (i) the one order-of-magnitude difference in Fa contents between type I and type II chondrules in the least metamorphosed chondrites may be seen e.g. in the plots by Ushikubo and Kimura (2021). (ii) As may be seen in the Fa histograms compiled by Brearley and Jones (1998), type II chondrule olivines equilibrate more rapidly with matrix olivine, because they were compositionally closer to it originally and ferroan olivine has larger diffusion coefficients (Chakraborty 2010). As such, with matrix+type II chondrules typically dominating the olivine budget together (chiefly matrix for carbonaceous chondrites and chiefly type II chondrules for ordinary chondrites), $Fa_{II}$ rapidly approaches the whole-rock value.

For such type II chondrule-bearing chondrites, we may then supersede the ideal formula with the better defined (if slightly larger) expression:

$$m = \frac{Fa_I}{Fa_{II}}$$

---

[8] We mean the olivine volume-weighted mean (which corresponds to the area-weighted mean for a representative section). Since the molar volume of olivine differs little across the isomorphic series between fayalite and forsterite, this corresponds to the fayalite content of an hypothetical mix of all (separated) olivines (of each chondrule type) homogenized in closed system. The volume-weighted mean is independent of the spatial resolution of the analyses (unlike the standard deviation).

(Yet we will have to return to the ideal formula for reduced chondrites)

Following our default rounding rule, the transition to type 4 (i.e. M1), i.e. olivine homogenization, could be redefined as $Fa_I/Fa_{II} \geq 0.95$. We note that this ratio would be relatively robust to interoperator (or electron microprobe) bias as any bias in favor e.g. of large olivines would in general lower both the numerator and the denominator (given normal zoning, whether igneous or diffusion-induced). Hence the limited differences (comparable to statistical errors) between literature values and ours for matched meteorites (except for the Scott and Jones (1990) data for CO3.8 Isna; Fig. 4). Supplement B shows that for a given number of analyses, evaluation of *m* is a more precise proxy of metamorphism than a *RSD(Fa)* estimated from a Dodd et al. (1967)-style data collection, simply because the former allows to preferentially sample type I chondrule olivines which are chiefly affected by equilibration.

It is worth cautioning that a given value of *m* does not correspond to a fixed time-temperature history for different parent bodies. Indeed, in the same way the PSF underlying the aqueous grades are affected by the abundance of fine-grained matrix, the equilibration of chondrule olivine depends on the crystal size distribution in chondrules. For a given time-temperature history, chondrites with smaller chondrules will thus acquire higher metamorphic grades. Metamorphic grades rather relate to the *ratio* between the diffusion length and a typical chondrule phenocryst size (see Supplement D).

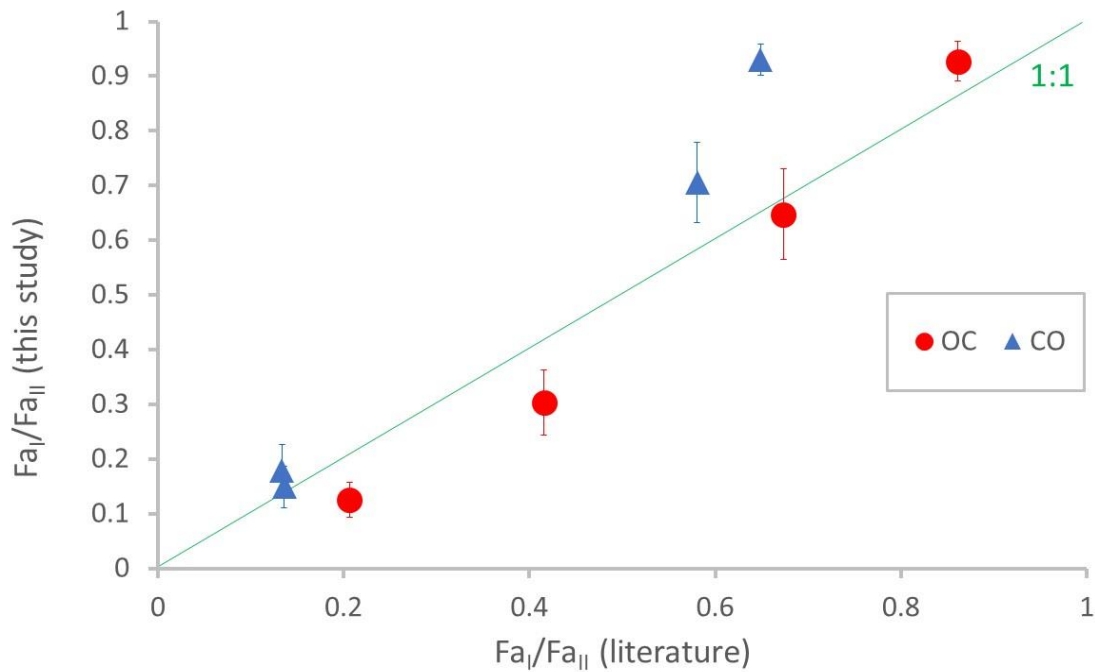

**Figure 4**: Comparison of $Fa_I/Fa_{II}$ measured in this study and literature values for matched meteorites (Scott and Jones 1990; McCoy et al. 1991; Noonan 1975; Noonan et al. 1977 (identifying in the latter Saint Mary's County "metal-rich" and "metal-poor" chondrules to type I and II, respectively)). Error bars are 1 sigma.

Although any $Fa_I/Fa_{II}$ value above 0.1 is the result of metamorphism and thus can serve as a measure thereof, the value for the least metamorphosed chondrites is not zero but depends on the pre-metamorphic chondrule populations. In CO chondrites, this is 0.03 (for ALH 77307; Scott and Jones 1990) and for UOCs 0.06 (for Semarkona; McCoy et al. 1991). We obviously cannot blindly apply the default rounding to the *nearest* tenth there, otherwise the OCs would start at M0.1. Also, even for CO, any interoperator bias in the measurement of $Fa_I$ would unduly affect the assessment of pristinity of the lowest-subtype samples, so fayalite content may not be the best assignation criterion for them.

For these lowest subtypes, however, chondrule olivine notoriously retains a sensitive tracer of metamorphism in the Cr content of the type II chondrule olivine (Fig. 5). Cr indeed is lost early from type II chondrule olivine, presumably to join chromite (Bunch et al. 1967), and this Grossman and Brearley (2005) used to refine the lowest subtypes of UOCs. Rubin and Li (2019) later proposed specific boundaries for CO chondrites. So we should adapt the rounding of *m* according to the mean $Cr_2O_3$ content of the type II chondrules, following these precedents, which have hitherto shown boundaries only in graphical form.

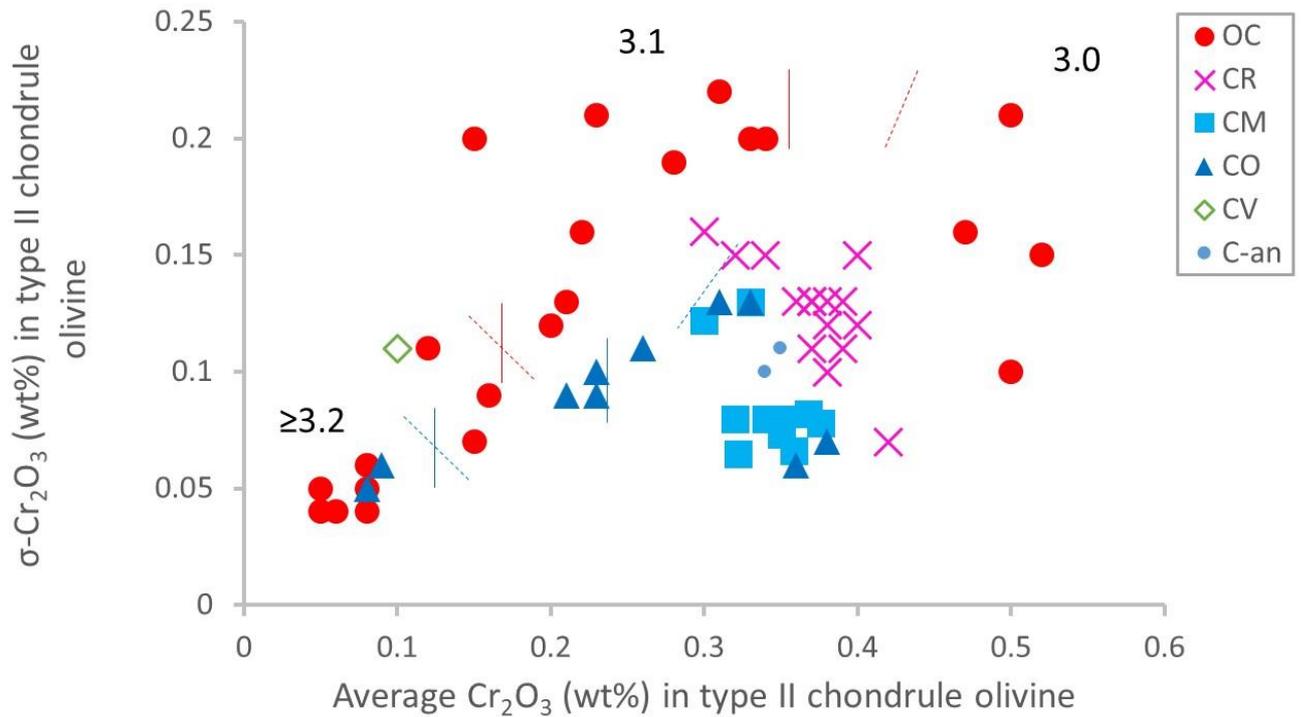

**Figure 5**: $Cr_2O_3$ average vs. standard deviation in type II chondrule olivine in unequilibrated chondrites. Data from Grossman and Brearley (2005), Schrader and Davidson (2017), Hewins et al. (2014), Schrader et al. (2015). Dashed lines represent the 3.05/3.1 and 3.15/3.2 boundaries proposed for UOC and CO by Grossman and Brearley (2005) and Rubin and Li (2019), respectively, and solid lines are our revisions thereof.

Since the starting $Cr_2O_3$ contents in type II chondrules was not the same in different chondrite groups, we cannot merely specify boundaries as a function of the absolute $Cr_2O_3$ content (as in Kimura et al. 2024) but we may reason in terms of its *relative* decrease since the onset of metamorphism. In fact, the 3.1/3.2 boundaries proposed by Grossman and Brearley (2005) for UOC and Rubin and Li (2019) for CO happen to correspond both to $Cr_2O_3$ at one third the highest measured (presumably pre-metamorphic) value $Cr_2O_3(0)$. Since this corresponds to *m* ~ 0.10-0.18 (Fig. 6), we may likewise accept $Cr_2O_3/Cr_2O_3(0)$ = 1/3 as the boundary between M0.1 and M0.2. (Supplement D also explores why, theoretically, there should be a (rough) link between $Cr_2O_3/Cr_2O_3(0)$ and *m*). In other words, M0.0 and M0.1 correspond to the loss of two thirds of type II chondrule olivine Cr. We then propose the M0.0/M0.1 boundary to mark the loss of the first third, i.e. a critical $Cr_2O_3/Cr_2O_3(0) = 2/3$ (this coincides with the 3.0/3.1 boundary of Kimura et al. (2024) in the case of CO chondrites). This would reproduce all Grossman and Brearley (2005) and most Rubin and Li (2019) classifications. Still, their graphs would put their 3.0/3.1 boundaries somewhat earlier (around $Cr_2O_3/Cr_2O_3(0) = 0.8$)[9] but these would increase the risk of really pristine chondrites being misclassified as M0.1 because of statistical or systematic errors in mean composition estimates[10]. Similar to Sears (2016), we *a fortiori* refrain from proposing finer subdivisions (e.g. the subtypes 3.00 and 3.05 of Grossman and Brearley (2005) or Rubin and Li (2019)), although one might base some on other criteria for specific groups (e.g. sulfide redistribution in matrix; Grossman and Brearley 2005). However finely defined, each subtype, including M0.0, must represent a finite range of metamorphic alterations rather than a discrete step.

---

[9] Strictly speaking, the oblique boundaries in these figures (and Fig. 5 herein) suggest that the subtypes should also depend on the standard deviation of $Cr_2O_3$ but we do not keep the latter as an assignation criterion as it critically depends on the exact data collection procedure (the lack of univariant trend among CO with additional data (e.g. Haenecour et al. 2018; Righter et al. 2024) suggests some variation among classifiers, possibly simply due to the variable importance of isolated olivine grains and repeat measurements). We may imagine that a real average (of all olivine grains) will be practically measurable someday by some automated mean (rather than the typical grains here). This would yield essentially the same value (see Supplement A) but with a distinct standard deviation.

[10] Also, if that threshold value of 0.8 were adopted, the reformulation in terms of $Cr_2O_3/MnO$ (with a threshold of 0.8 × 1.2 = 0.96) in the succeeding paragraph would put nearly all CRs at M0.1 and the all-pristine Acfer 094 (1.03 ± 0.06; Grossman and Brearley 2005) within 1 sigma of it.

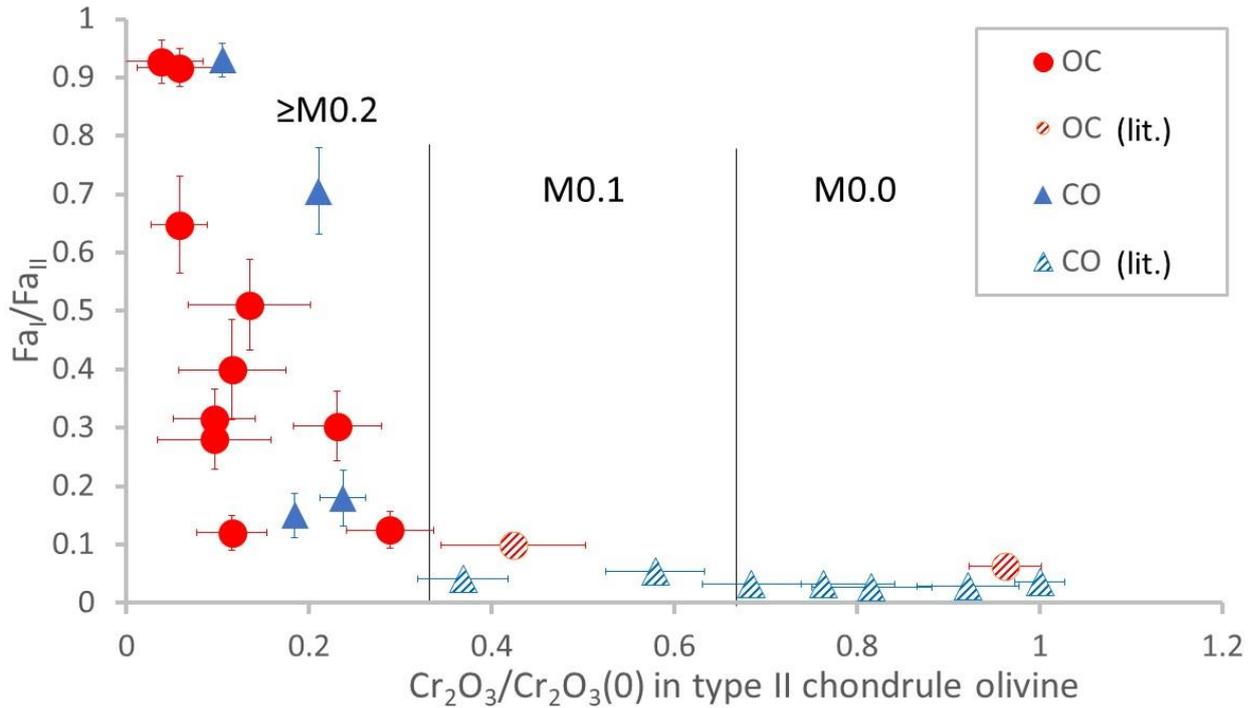

**Figure 6**: $Fa_I/Fa_{II}$ as a function of type II chondrule olivine $Cr_2O_3$ normalized to the pre-metamorphic value in UOC and CO chondrites. Vertical lines indicate our boundaries for M0.0, M0.1 and M0.2 (i.e. subtypes 3.0, 3.1 and 3.2 for type 3 chondrites). Cr data from Grossman and Brearley (2005), MetBullDB (for CO) and this study; literature Fa data from McCoy et al. (1991) for Semarkona and Jacquet et al. (2015) for Bishunpur. Error bars are 1 sigma.

  Now, this redefinition of the lowest subtypes presupposes knowledge of the initial value $Cr_2O_3(0)$, which would again depend on the identification of the chemical group. Yet MnO seems to provide a suitable proxy, in that the most unmetamorphosed chondrites generally show $Cr_2O_3/MnO \approx 1.2$ in their type II chondrule olivine (Fig. 7). Why should that be? Bulk type II chondrules have nearly chondritic $Cr_2O_3/MnO$ ratios (e.g. Gordon 2009) given the close volatilities of Mn and Cr (half-condensation temperatures of 1158 vs. 1296 K; Lodders 2003) and the limited devolatilization of such chondrules (Gordon 2009). Alexander and Ebel (2012) report an average type II bulk ratio of 1.3 (compared to 1.5 for CI; Lodders 2003). Since the olivine/melt partition coefficients are close for both elements (Jing et al. 2024), the fractional crystallization of olivine should not incur much change in that ratio in olivine. This is supported by the relatively parallel zoning profiles of Mn and Cr in olivine (notwithstanding the final drop in $Cr_2O_3$ due to incipient chromite precipitation; Jones 1990) and the overall correlation between $Cr_2O_3$ and MnO in both Semarkona (Jones 1994) and ALH 77307 (Jones 1992) type II chondrule olivine. Whatever the value of this explanation, we may thus set threshold $Cr_2O_3$/MnO values of $1.2 \times 2/3 = 0.8$ and $1.2/3 = 0.4$ for the M0.0/M0.1 and M0.1/M0.2 boundaries, respectively.

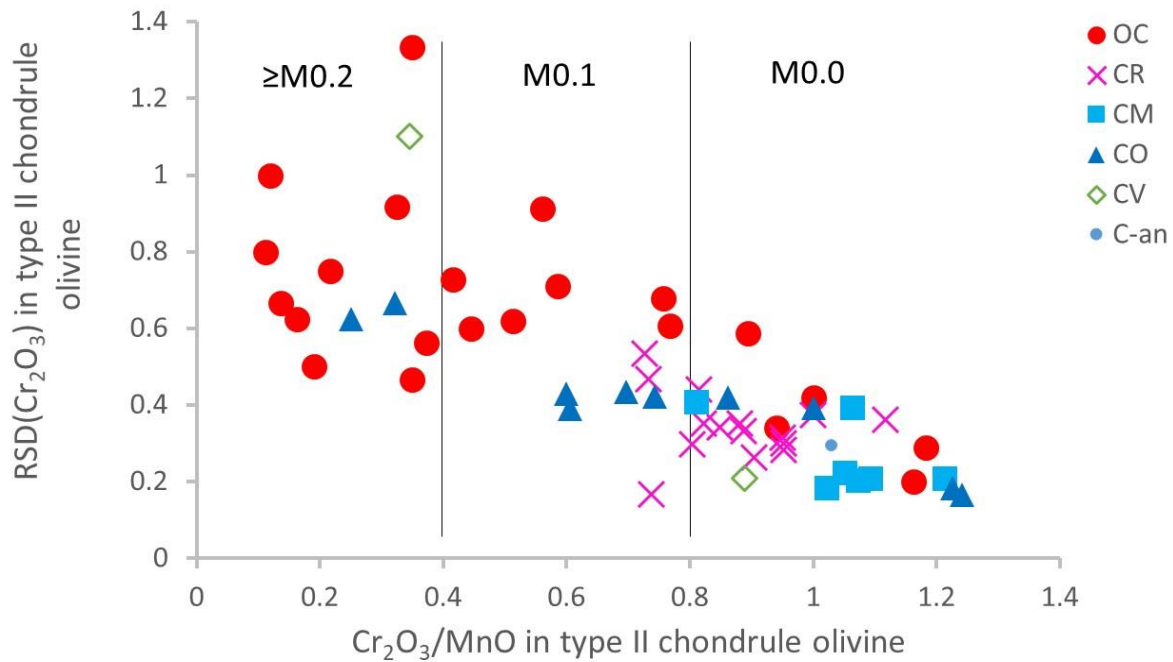

**Figure 7**: Relative standard deviation of $Cr_2O_3$ as a function of $Cr_2O_3$/MnO (ratio of the means) in type II chondrule olivine. Same data sources as Fig. 5. Vertical lines indicate our proposed boundaries for M0.0, M0.1, M0.2.

However lengthy the justification above, the resulting classification algorithm (for type II chondrule-bearing unequilibrated chondrites) is simple and may thus be recapitulated as follows:

1° Measure type I and type II chondrule olivine compositions (one typical grain of the closest olivine-dominated chondrule per random point; see the Method section)

2° Calculate mean type II olivine composition:

If $Cr_2O_3$/MnO $\geq 0.8$, metamorphic grade is M0.0.

> If $0.4 \leq Cr_2O_3/MnO < 0.8$, metamorphic grade is M0.1.
>
> 3° Else, calculate mean type I Fa composition and the ratio $m = Fa_I/Fa_{II}$. The metamorphic grade is M$r(m)$ where $r(m)$ is $m$ rounded to the nearest tenth $\geq 0.2$.
>
> 4° Barring significant aqueous alteration, the petrologic subtype is $3+r(m)$.

Again, this only governs the subtype indicated by olivine composition. Other assignation criteria may overrule it, depending on the adaptations made for each chondrite group. These we now move on to consider.

### 5.1 Ordinary chondrites

We announced as our touchstone that our redefinition of subtypes should match closely those of Sears et al. (1980) for ordinary chondrites. We have yet to show that our proposed formula for $m$ fulfills this requirement.

By construction, this is already the case for all subtypes 3.0-3.1 (M0.0-M0.1) as recalibrated by Grossman and Brearley (2005). Strictly speaking, the reformulation in terms of $Cr_2O_3/MnO$ fails for H3.10 chondrite RC 075, which shows a $Cr_2O_3/MnO = 0.89 > 0.8$, but this is due to an anomalously low mean MnO of 0.38 wt% (±0.21). If the highest $Cr_2O_3$ of the Grossman and Brearley (2005) study is taken as a reference for the initial $Cr_2O_3$ (0.52 ± 0.15 wt% for QUE 97008) the M0.0/M0.1 and M0.1/M0.2 boundaries in terms of $Cr_2O_3$ can be set at ~0.35 and 0.17 wt% $Cr_2O_3$, respectively, matching the Grossman and Brearley (2005) classifications without exception (Fig. 5).

For more metamorphosed UOCs, we see from Fig. 8, that $m$ generally matches the decimal part of the literature subtype within errors (~one decimal subtype), but there are clear outliers which we discuss now.

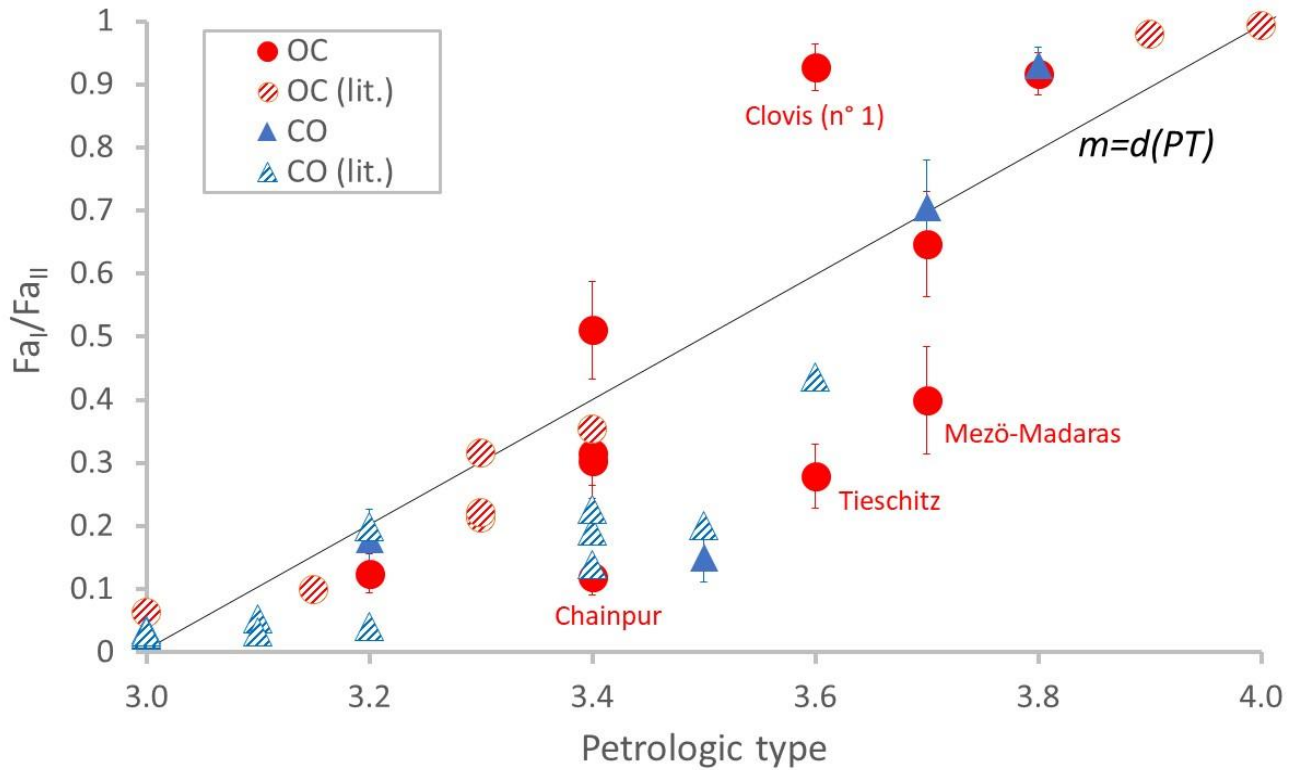

**Figure 8**: Fa$_I$/Fa$_{II}$ as a function of petrologic subtype for UOC and CO chondrites. For UOCs, data scatter along the diagonal line corresponding to m equaling the decimal part of the petrologic type (PT), with named outliers discussed in the text. For CO, as will be seen in their dedicated subsection, the interval 3.3-3.6, which deviates most clearly from the 1:1 line, is hitherto ill-defined and requires revision. Literature data from McCoy et al. (1991), Jacquet et al. (2015), Baeza (2017), DeHart (1989), Noonan et al. (1977), Scott and Jones (1990), MetBullDB (for CO).

Mezö-Madaras is classified as L3.7 (e.g. Sears et al. 1980) but is here ranked as M0.4. The discrepancy is a clear consequence of its genomict character, as equilibrated L4-5 clasts have long been described in this meteorite (e.g. Binns 1968; Fig. 9). Since TL sensitivity gains one order of magnitude every three subtypes, ~10 % of equilibrated clasts in an otherwise low-subtype host suffices to bring a bulk TL sensitivity at subtype 3.7 level. The same explanation seems to hold for Chainpur (hitherto LL3.4 but M0.2): here, only ~1 % of equilibrated material would be needed, and this is borne out by the occasional observation of quite equilibrated isolated chondrules (not part of larger clasts): two (out of 114 chondrules examined) with crystalline feldspar and olivine Fa consistent with LL4 chondrites interpreted by Ruzicka (1990) to have been annealed before agglomeration, or two type I chondrules (#4 and 5 in fraction 4) out of 24 studied by Keil et al. (1964) with likewise LL4-like olivine Fa contents but Fs around 10 mol%. We also observed such chondrules in our section, one of which is depicted in Fig. 1b.

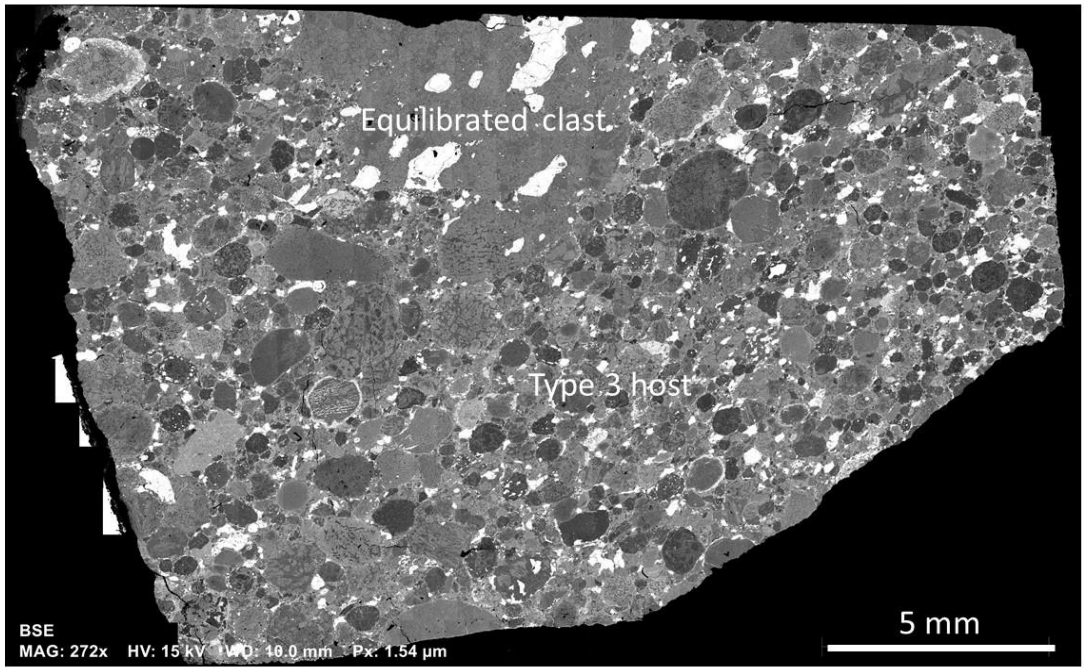

**Figure 9**: Back-scattered electron map of section 4416sp1 of ordinary chondrite Mezö-Madaras (hitherto L3.7 but reclassified here as L3.4). An equilibrated clast is visible on the middle top.

Thus, the strong dependence of TL sensitivity on metamorphism, a strength of the Sears et al. (1980) classification, also presents a drawback in incurring a strong sensitivity to equilibrated xenoliths. This is nonnegligible as most type 3 ordinary chondrites are brecciated at thin section scale (in particular 86 % of LL3s; Bischoff et al. 2018), although only a fraction of those breccias may be genomict. In this case, the bulk TL value may correspond to no lithology (as in Mezö-Madaras and Chainpur), let alone the host (which would give the maximum metamorphic level of the meteorite as an ensemble). In contrast, $Fa_I/Fa_{II}$ of a genomict ensemble would roughly correspond to the volume-weighted average of the $Fa_I/Fa_{II}$ of the different lithologies[11]. Thus the effect of minor lithologies would be minimal, and can of course be further mitigated by avoiding them in the point selection (as we have done here for Mezö-Madaras). Lithologies comprising at least 10 vol% (Jacquet 2022) can have their metamorphic degrees evaluated independently so as to display the full range of petrologic types in the hyphenated way customary of MetBull classifications. So Mezö-Madaras can be described as L3.4-4 or simply L3.4 if only the host subtype is meant[12]; the full information could be shown by writing "L3.4 (3.4-4)".

---

[11] Strictly speaking, this would be the average of that ratio *multiplied* by the bulk-normalized modal ratio of type I to type II olivine of the associated lithology, and *weighted* by the type II-hosted number of fayalite moles.

Another outlier is Tieschitz, hitherto classified as type 3.6 (Sears et al. 1980) but ranked as M0.3 here. Here the TL sensitivity has been likely increased by the hydrothermal deposition of "white matrix" (Christophe Michel-Levy 1976) which Dobrica and Brearley (2014) showed to be polycrystalline albitic plagioclase. Its classification as 3.6 previously conflicted with limited matrix recrystallization or FeO depletion (Sears et al. 1982). Still, the low Co variability in kamacite, low bulk C or $^{36}$Ar contents seemed consistent with higher subtypes (3.8, 3.7 and 3.6 respectively; Sears et al. 1982), but from the plots of Sears et al. (1980), Co variability actually has little discriminatory power above subtype 3.2 (and at the time was only evaluated for LLs below), C likewise little below 3.8, and $^{36}$Ar merely sets Tieschitz beyond Krymka (LL3.2), Chainpur, Mezö-Madaras, and before Manych (LL3.4; M0.3), and Parnallee (LL3.7; M0.7), i.e. mostly "low-subtype" UOCs.

A final outlier, Clovis (n° 1), classified as H3.6 by Sears et al. (1982) appears here *more* metamorphosed (M0.9) than its TL data suggested. Data by Noonan (1975) also indicate $m = 0.86$ if his "distinct chondrules" and "angular and vesicular fragments" are interpreted as type I and II chondrules respectively judging from his photomicrographs. In fact, Sears et al. (1980) themselves had originally predicted a 3.9 classification based on olivine PMD. The lack of $Fe^{2+}$ disorder in pyroxene, consistent with equilibrated chondrites (Dundan and Walter 1967), and the olivine-spinel closure temperature of 710±40 °C, just above H3.8 chondrite Dhajala (Wight et al. 1994) are also consistent with such a high subtype. One may surmise that shock amorphized feldspar, as "glass" is reported by Noonan (1975), but we do not know the shock classification of

---

If the clasts have the same type I and type II chondrule modal abundances, the second factor is unity, and if Fa$_{II}$ does not vary much with metamorphism the weight coefficient is proportional to the volume fraction of the lithology.

[12] Had we preferred a "bulk" subtype representative of the average of all lithologies, it would have been 3.5 (for the 10 % equilibrated material estimated above for the whole-rock), within error of the (more precisely evaluable) host subtype though.

Clovis (n° 1). Only 3 out of 111 H3 chondrites studied by Bischoff et al. (2019) reached the shock stage S4 corresponding to partial maskelynitization, and only one H3 studied by Miyahara et al. (2020) had shock-induced melting textures. So although definitely a possibility here, shock amorphization is probably a minor disturbance for TL-based UOC classification in general.

Thus, the $m$-based subtypes seem to capture the essence of the Sears et al. (1980) subtypes (in a manner more robust against impact-related or other disturbances). Table 4 keeps essentially all Sears et al. (1980) criteria unchanged[13] but replaces their olivine column by the $m$ criterion. We stress again that revision of the Sears et al. (1982) olivine compositional criteria *was* necessary (Fig. 3) independently of our goal of intergroup harmonization.

| Metamorphic grade | M0.0 | M0.1 | M0.2 | M0.3 | M0.4 | M0.5 | M0.6 | M0.7 | M0.8 | M0.9 |
|---|---|---|---|---|---|---|---|---|---|---|
| Corresponding Sears et al. (1980) subtype | 3.0 | 3.1 | 3.2 | 3.3 | 3.4 | 3.5 | 3.6 | 3.7 | 3.8 | 3.9 |
| Example | Semarkona | Bishunpur | Krymka | Tieschitz | Sharps | ALH 77011 | Khohar | Parnallee | Dhajala | Clovis (n°1) |
| $TL_{Dhajala}$ | <0.0046 | 0.0046-0.01 | 0.01-0.022 | 0.022-0.046 | 0.046-0.1 | 0.1-0.22 | 0.22-0.46 | 0.46-1 | 1-2.2 | >2.2 |
| Type II chondrule olivine $Cr_2O_3$ (wt%) | >0.35 | 0.17-0.35 | <0.17 | <0.17 | <0.17 | <0.17 | <0.17 | <0.17 | <0.17 | <0.17 |
| $Fa_i/Fa_{II}$ | <0.1 | 0.05-0.15 | 0.15-0.25 | 0.25-0.35 | 0.35-0.45 | 0.45-0.55 | 0.55-0.65 | 0.65-0.75 | 0.75-0.85 | 0.85-0.95 |
| F/FM | >1.9 | 1.7-1.9 | 1.6-1.7 | 1.5-1.6 | 1.4-1.5 | 1.3-1.4 | 1.2-1.3 | 1.1-1.2 | 1.0-1.1 | ≥1.0 |
| Matrix recrystallization (vol%) | <10 | 10-20 | 10-20 | 10-20 | ~20 | ~50 | >60 | >60 | >60 | >60 |
| C (wt%) | ≥0.6 | 0.5-0.6 | 0.43-0.5 | 0.38-0.43 | 0.33-0.38 | 0.3-0.33 | 0.27-0.30 | 0.24-0.27 | 0.21-0.24 | ≤0.21 |
| $^{36}Ar$ ($10^{-13}$ $m^3$STP/kg) | ≥65 | 55-65 | 45-55 | 35-45 | 27-35 | 18-27 | 13-18 | 8-13 | 4-8 | <4 |
| $I_D/I_G$ | <0.9 | 0.9-1 | >1 | >1 | >1 | >1 | >1 | >1.6 | >1.6 | >1.6 |
| $FWHM_D$ ($cm^{-1}$) | >170 | 140-170 | <140 | <140 | <140 | <140 | <140 | <70 | <70 | <70 |

**Table 4**: Metamorphic scale for UOCs. Most rows are copied *verbatim* from Sears et al. (1980), with "F/FM" is the matrix FeO/(FeO+MgO) normalized to the whole-rock value and $TL_{Dhajala}$ the induced thermoluminescence sensitivity normalized to Dhakala. We have however replaced the olivine criteria and added Raman spectroscopic criteria following the works of Quirico et al. (2003) and Bonal et al. (2006, 2007), with FWHM-D being the full width at half maximum of the D band and $I_{D,G}$ the peak intensity of the D, G bands.

However limited our modification of the Sears et al. (1980) classification is then, wherever $m$ clearly changes (beyond analytical error) the balance of the criteria (especially if contradicting ones can be explained away), the meteorite ought to be reclassified accordingly. All our above outliers should thus be reclassified as shown in Table 1. The reclassifications of Chainpur and Tieschitz are of some moment for the Raman-based classification, since they were Quirico et al. (2003) and Bonal et al. (2006, 2007)'s reference samples for subtypes 3.4 and 3.6, respectively. So the Raman subtype assignation boundaries ought to be redefined—or rather defined, as in fact the aforementioned works refrained from setting explicit boundaries (although Righter et al. (2022) drew some in their Fig. 2). Fig. 10-11 leave open the possibility of a fair inter-group synchronization of *m*- and Raman-based classifications at early stages. We have thus proposed the same M0.0/M0.1 and M0.1/M0.2 Raman boundaries for OCs and CCs (Table 4-6), allowing a simultaneous use of organic matter maturity as a universal sensor of metamorphism at least until M0.2.

---

[13] We nevertheless elect to forget about the PMD of kamacite Co, because of its limited discriminatory power mentioned above, and also as part of a move to eradicate PMDs (as opposed to the more practical standard deviation) from meteorite taxonomy (e.g. Grossman 2011).

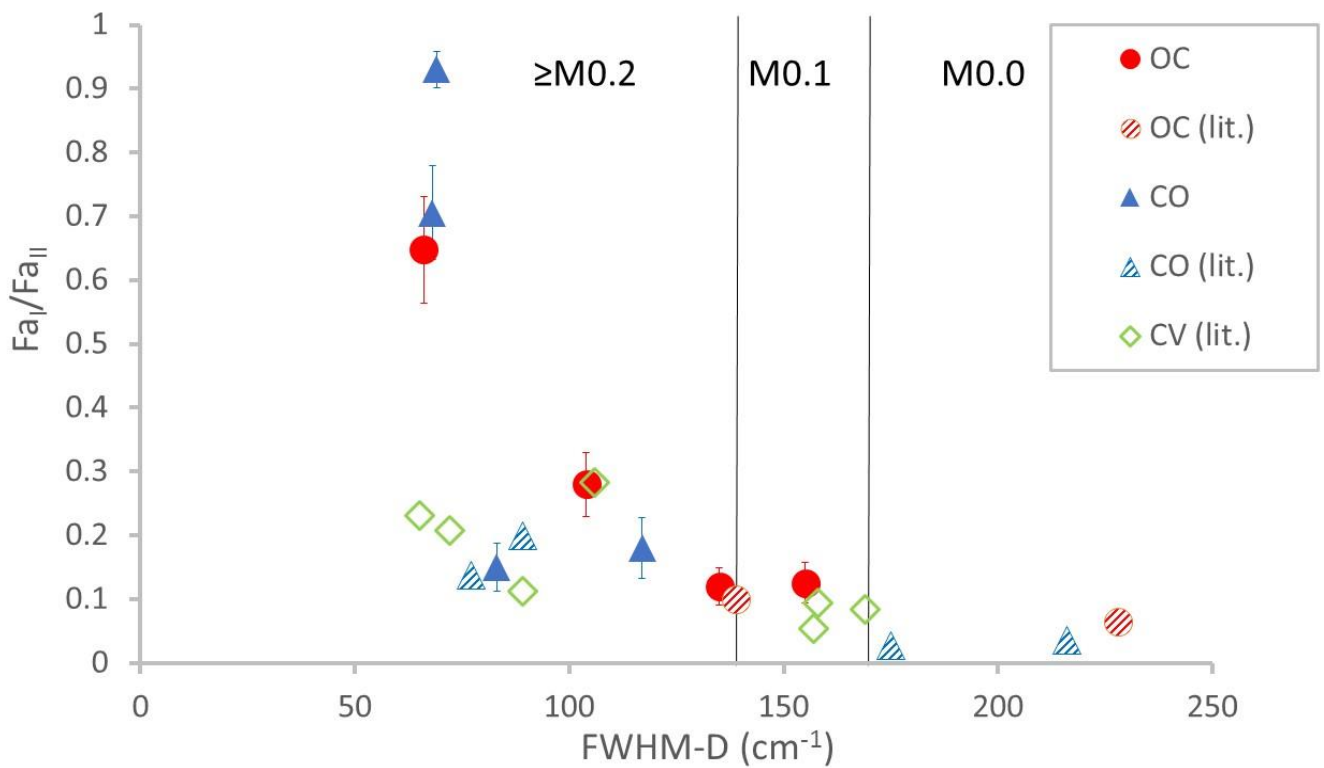

**Figure 10**: $Fa_I/Fa_{II}$ as a function of Full Width at Half Maximum of Band D (FWHM-D) of the Raman spectrum of organic matter, with proposed boundaries for M0.0, M0.1 and M0.2. Raman data from Quirico et al. (2003), Bonal et al. (2006, 2007); same Fa data sources as Fig. 8, plus Simon et al. (1995) for Axtell and Allende, Karcher et al. (1985) for Leoville, Hutchison & Symes (1972) for Vigarano, Van Schmus (1969) for Grosnaja and Kaba, Wood (1967) for Mokoia. For all these CVs, $Fa_I$ was taken to be the average of the data below $Fa_{20}$ (that is, the hiatus preceding type II values).

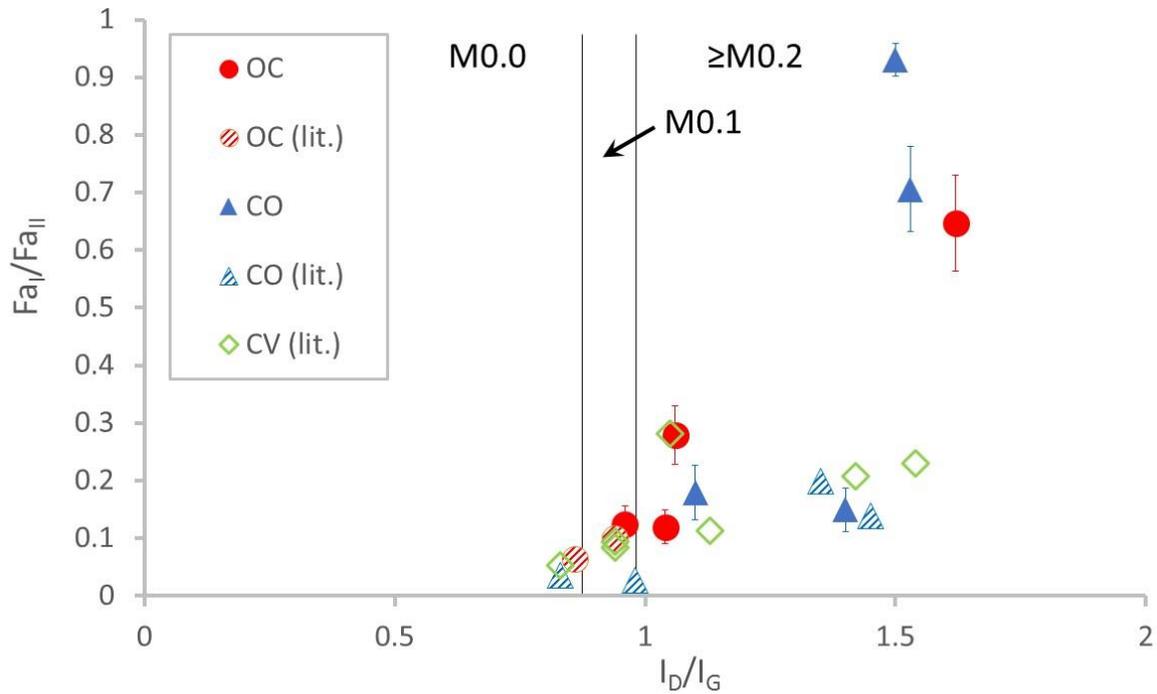

**Figure 11**: Fa$_I$/Fa$_{II}$ as a function of the ratio of the peak intensities of the D and G bands of the Raman spectrum of organic matter. Same data sources as Fig. 10.

### 5.2 Rumuruti chondrites

Bischoff (2000) inaugurated the practice of assigning subtypes to R3 chondrites using the *RSD(Fa)* formula of Sears et al. (1982). The relationship between *m* and *RSD(Fa)* depends on $x_I$. To estimate it, we note that, in weakly metamorphosed R chondrites, Friend et al. (2017) analyzed 227 low-Ca pyroxenes below Fs$_{10}$ and 106 beyond, hence a pyroxene type I/type II ratio of $x_{I,Px}/(1-x_{I,Px}) = 227/106 = 2.1$ (where $x_{I,Px}$ is the fraction of chondrule pyroxene hosted by type I chondrules), which should be representative of the pre-metamorphic ratio given the slow Fe-Mg interdiffusion in pyroxene compared to olivine. If, owing to the likely genetic link with ordinary chondrite chondrules (e.g. Regnault et al. 2022), we apply the same *θ* defined in Supplement E.2, we can infer $x_I = 0.6$.

From Fig. 12, such a value would constrain *m* to be close to 1-*RSD(Fa)*, i.e. the decimal part of the Bischoff et al. (2000) subtype. Thus our scheme would likely not incur marked changes to the current practice. Yet a dedicated study on R chondrite metamorphic classification, including direct measurements of *m* and other parameters, would be worthwhile.

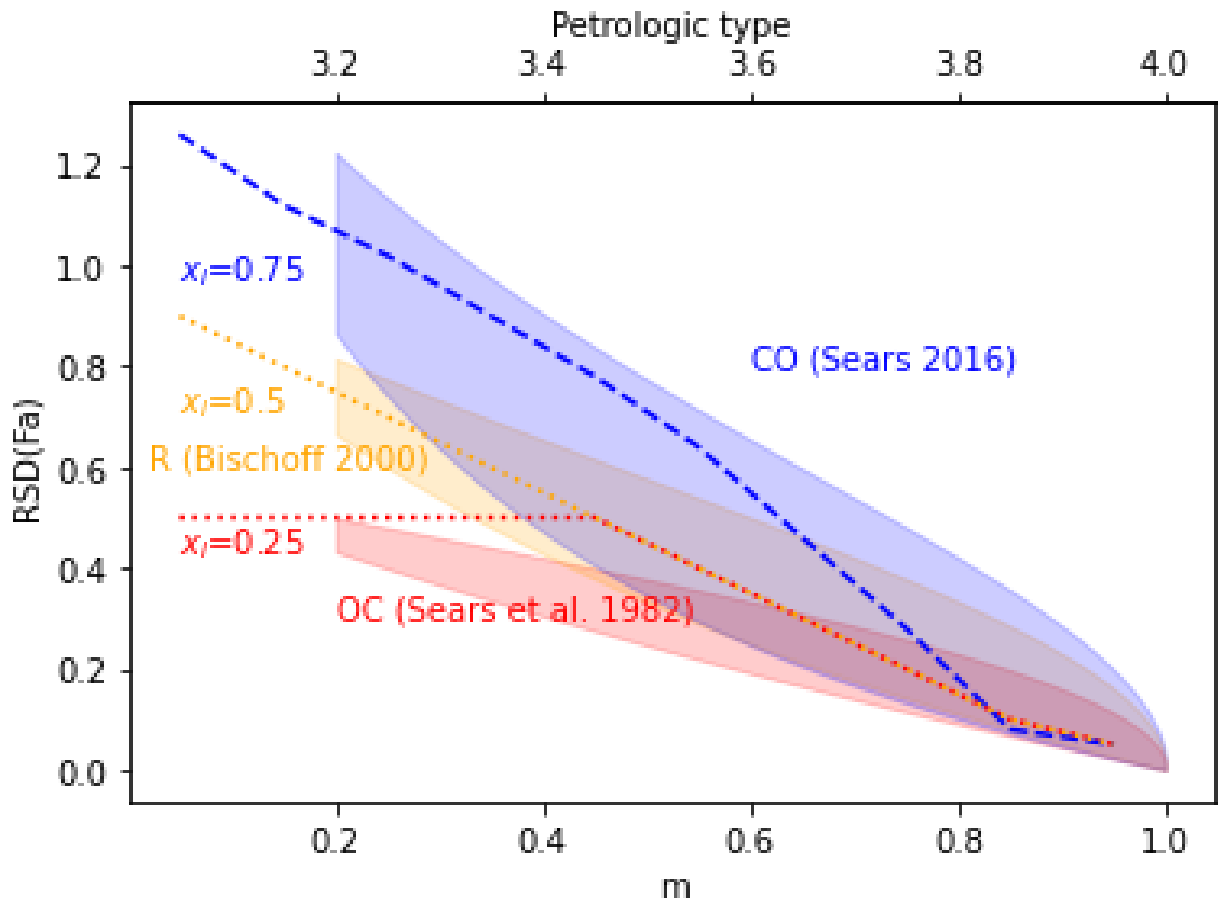

**Figure 12**: Theoretical relationship (shaded areas) between *m* and *RSD(Fa)* for different values of $x_I$ (for $m \geq 0.2$). See Supplement C.2 for justification. Also overplotted are the relationships between petrologic subtype and *RSD(Fa)* for different literature schemes, as in Fig. 3.

5.3 CO chondrites

The current MetBullDB CO subtypes mainly follow Chizmadia et al. (2002), who proposed a classification of CO chondrites based on the introduction of Fe in amoeboid olivine aggregates (AOA). The interval 3.0-3.2 was further refined by Rubin and Li (2019). Other schemes include Scott and Jones (1990) and Sears (2016). How does our metamorphic scale compare with the literature?

By design, the $Cr_2O_3$-based definitions of M0.0 and M0.1 match the literature usage for CO3.0 and CO3.1. From MetBullDB data, only NWA 6701, NWA 6447 and Los Vientos 123 would have to be reclassified from CO3.1 to CO3.0 (=M0.0); and NWA 8345 (currently CO3.2) would (narrowly) become a CO3.1. We note that in refraining from refining subtypes as in Rubin and Li (2019), we avoid assigning different samples from the same strewnfield to three different subtypes. Indeed Righter et al. (2024) have their dominant Transantarctic mountain pairing groups both straddle Rubin and Li (2019)'s subtypes 3.05-3.15, while they fit entirely in our M0.0.

Nominally, the higher subtypes also match with the literature. Although only Scott and Jones (1990) provided indicative values for $Fa_I$, they and Sears (2016) gave criteria in terms of the average Fa content $\overline{Fa}$. Fig. 13 shows that they closely overlap with ours. The $RSD(Fa)$ thresholds should also roughly match with our classification (Fig. 12).

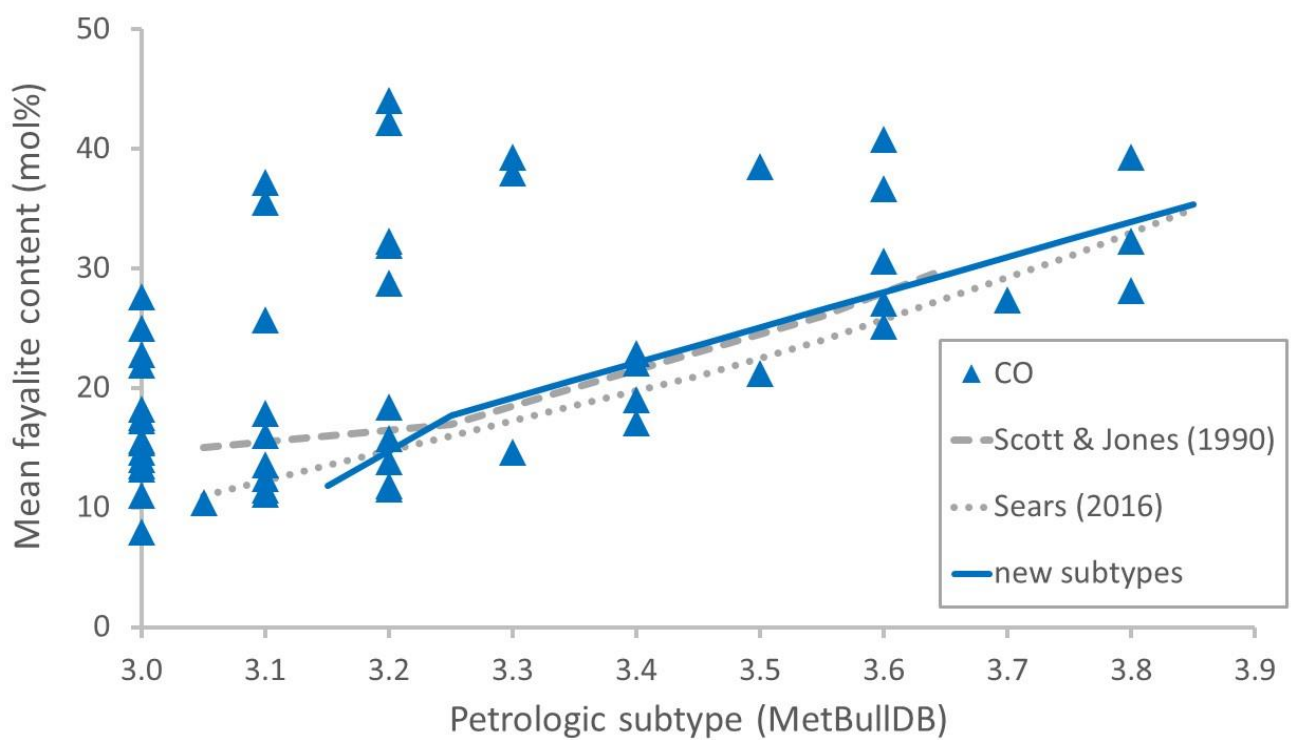

**Figure 13**: Average fayalite content of CO as a function of petrologic subtype. Overplotted are the values for the Scott and Jones (1990) and Sears (2016) subtypes, as well as ours if one assumes $Fa_{II}$ = 40 mol% (the average of our four samples) beyond 3.2 and $x_I$ = 0.74 (see Supplement C.1).

Now, in practice, MetBullDB data scatter significantly away from the nominal curves of either scheme, and it is often unclear which criterion was used for the classification. Yet most MetBullDB entries which give a reason for a subtype do invoke olivine composition (with Raman spectra also used for the lowest subtypes). At any rate, the "average" Fa depends on the selection thresholds for olivine grains (e.g. when would an isolated grain in the matrix count? etc.) and is thus of questionable reliability.

The disagreement is especially strong for literature subtypes in the range 3.3-3.6. In particular, the "classical" CO chondrites making McSween (1977)'s stages I and II—Kainsaz, Felix, Ornans, Lancé—are found here to belong all to the M0.2 grade. We find no significant difference in terms of the Fa content of the type I chondrules between the extremes Kainsaz and Lancé (CO3.2 and CO3.5 in MetBullDB respectively). To be sure, the oft-cited Van Schmus (1969) dataset reported $\overline{Fa} = 11.8$ mol% for Kainsaz contrasting with $\overline{Fa} = 18.4 - 21.2$ mol% for Felix, Ornans and Lancé. However, the associated histograms (see also Crabb et al. (1982)) make plain that the mean Fa for $Fa_{<20}$ olivine (essentially the type I population, given the observed hiatuses around $Fa_{20}$) in Kainsaz and Lancé are indistinguishable (6±4 vs. 5±5 mol%; 1σ) and close to our $Fa_I$ values. The difference in the overall mean Fa values lies in the lower abundance of more ferroan olivine grains selected in the Van Schmus (1969) Kainsaz section (12 % vs. 38 % in Ornans and Lancé in the same work, again assuming the $Fa_{20}$ cutoff).

While Bonal et al. (2007) already ranked all these chondrites at the same subtype (their 3.6), Sears (2016) still set Kainsaz (which made all McSween (1977)'s stage I) apart (at 3.2 vs. 3.4). Now, disregarding olivine, Kainsaz still appears more "primitive" than all McSween (1977) stage II chondrites for 5 out of 10 Sears et al. (2016) criteria (kamacite composition, $^{36}$Ar content, organic matter maturity) even though its TL sensitivity is higher than all of them. Kainsaz may thus still be less metamorphosed in some sense than other M0.2 CO chondrites, but this does not necessarily warrant a distinct subtype. For one, as we just saw, not all criteria give the same order so further refinement below the scale of fluctuations (analytical or real) off an assumed monotonous trend may be excessive. Also, setting Kainsaz apart could make sense at the time of McSween (1977) where it appeared the most primitive CO known but since many CO3.0 and CO3.1 have been recognized afterward, any possibly useful refinement should be restricted to them (e.g. Rubin and Li 2019).

Table 5 lists our metamorphic grades as adapted for CO chondrites. We did not retain among our criteria the thickness of ferroan olivine veins in AOAs reported by Chizmadia et al. (2002) because they may have precipitated from fluids (e.g. Han et al. 2022) and thus not directly trace metamorphism after all. Given that many "classical" CO chondrites are lumped together as M0.2, while the McSween (1977) "stage III" chondrites Warrenton and Isna retain their literature subtype (even though we measured M0.9 for Isna, but within errors of its 3.8 classification), the hereby depopulated interval M0.3-M0.6 suffers from a paucity of well-characterized chondrites. Yet there is no real hiatus, as suggested by Fig. 13. We assigned ALH 77003 to M0.4 (a moderate correction to Scott and Jones (1990)'s subtype 3.5) given the ratio $Fa_I/Fa_{II} = 0.44$ found by Scott and Jones (1990) and independently confirmed by the formula of supplement C.1 applied on their randomized histogram, using $x_I$=0.61 (extrapolated from their ALH 77307 and ALH 82101 data) and their $Fa_{II}$ value (this yields $Fa_I/Fa_{II} = 0.42$). This grade and other "intermediate" metamorphic grades cannot at present be better characterized for non-olivine criteria. A dedicated study of the more metamorphosed CO would be worthwhile.

Among chondrites currently classified as CO, a matrix-poor "CA" grouplet has been set apart by Kimura et al. (2022), with an assigned subtype of 3.2. Cr systematics of ferroan olivine and Raman properties are indeed consistent with a M0.2 classification, but may marginally suggest M0.1 for Asuka-9003 and Asuka-09535.

| Metamorphic grade | M0.0 | M0.1 | M0.2 | M0.3 | M0.4 | M0.5 | M0.6 | M0.7 | M0.8 |
|---|---|---|---|---|---|---|---|---|---|
| Corresponding classification | CO3.0 | CO3.1 | CO3.2 | CO3.3 | CO3.4 | CO3.5 | CO3.6 | CO3.7 | CO3.8 |
| Example | ALH 77307 | MIL 090010 | Ornans | ? | ALH 77003 | ? | ? | Warrenton | Isna |
| McSween (1977) stage | - | - | I, II | - | - | - | - | III | III |
| Scott & Jones (1990) subtype | 3.0 | 3.0 | 3.1-3.4 | - | 3.5 | - | - | 3.6 | 3.7 |
| Chizmadia et al. (2002) subtype | 3.0 | 3.1 | 3.2-3.5 | - | 3.6 | - | - | 3.7 | 3.8 |
| Bonal et al. (2007) subtype | 3.03-3.1 | 3.1 | 3.6, >3.6 | - | - | - | - | ≥3.7 | ≥3.7 |
| Sears (2016) subtype | 3.0 | 3.1 | 3.2-3.4 | - | - | - | - | 3.6 | 3.7 |
| Type II chondrule olivine $Cr_2O_3$ (wt%) | >0.25 | 0.12-0.25 | <0.12 | <0.12 | <0.12 | <0.12 | <0.12 | <0.12 | <0.12 |
| $Fa_I/Fa_{II}$ | <0.05 | 0.05-0.15 | 0.15-0.25 | 0.25-0.35 | 0.35-0.45 | 0.45-0.55 | 0.55-0.65 | 0.65-0.75 | >0.75 |
| Diffusive halo thickness in AOA (µm) | invisible | invisible | <1 | >1 | 2-3 | >3 | >3 | (6.5) | (6.5) |
| $I_D/I_G$ | <0.9 | 0.9-1 | 1-1.5 | >1.5 | >1.5 | >1.5 | >1.5 | (1.53) | (1.5) |
| $FWHM_D$ (cm$^{-1}$) | >170 | 140-170 | 80-140 | <80 | <80 | <80 | <80 | (68) | (69) |
| $TL_{Dhajala}$ | <0.017 | 0.017-0.030 | 0.030-0.17 | >0.17 | (0.34) | >0.3 | >0.3 | (0.4) | (1) |
| kamacite Co (wt%) | <0.4 | 0.4-0.6 | 0.6-1 | >1 | (1.4) | >1 | >1 | (1.5) | (1.6) |
| kamacite Cr (wt%) | >0.5 | 0.4-0.5 | 0.2-0.4 | <0.2 | (0.15) | <0.2 | <0.2 | (<0.1) | (<0.1) |

**Table 5**: Metamorphic scale for CO chondrites. Raman parameters as in Table 4 (Bonal et al. 2006), TL after Sears (2016), kamacite from Scott and Jones (1990).

### 5.4 CV chondrites

Many CVs show BSE-visible evidence of influx of Fe in type I chondrule olivine (e.g. Bonal et al. 2007), but calculated *m* values do not exceed 0.3 (e.g. Fig. 10-11) for mainstream CVs. We set apart the C3 chondrite NWA 8418 which MacPherson et al. (2023) classified as CV4. Since its matrix (probably close to type II chondrules) averages $Fa_{38}$ while chondrule olivines (which are type I in great majority) show $Fa_{20\pm14}$ (if the histogram in MacPherson et al. (2023)'s Fig. 15 is representative), it would belong to metamorphic grade M0.5 and be a CV3.5.

With most CVs not exceeding M0.2, type II chondrule olivine Cr systematics should be decisive in principle, yet type II chondrules are too rare in CV chondrites for this to be practical (Grossman and Brearley 2005). To reliably estimate *m* in Fig. 10-11, we have even had to adopt a constant $Fa_{II}$ = 30.3 mol%, the average of all Grossman and Brearley (2005) CV analyses, but with higher degrees of metamorphism, Fe-Mg exchange with more ferroan matrix should increase it somewhat (as in NWA 8418).

Raman properties of organic matter may then be the handiest tool for CV classification. Bonal et al. (2007) classified CVs according to the Raman spectra of their carbonaceous matter in three broad OC-inspired subtype ranges, i.e.

"3.1", "3.1-3.4" and "3.6" (which we lump with their ">3.6"). They roughly correspond to $m \sim 0.05$, 0.1 and 0.2 respectively (Fig. 10-11), so can be received as M0.0, M0.1 and M0.2, respectively[14]. Table 6 recapitulates the resulting metamorphic classification for CV chondrites. The subtypes are considerably coarser, and thus simplified, relative to Bonal et al. (2007) or Righter et al. (2024).

Our Bonal et al. (2007)-inspired (if simplified) subtypes correlate little with the TL-based ones (affected by CAIs?) of Guimon et al. (1995), even though the latter roughly spanned our suggested range of subtypes (3.0-3.3). So we can propose neither TL criteria nor meaningful correspondences with those subtypes.

| Metamorphic grade | M0.0 | M0.1 | M0.2 | M0.5 |
|---|---|---|---|---|
| Corresponding classification | CV3.0 | CV3.1 | CV3.2 | CV3.5 |
| Example | Kaba | Vigarano | Allende | NWA 8418 |
| Bonal et al. (2006) subtype | 3.1 | 3.1-3.4 | 3.6, >3.6 | - |
| $Fa_I$ (mol%) | <2 | 2-5 | 5-8 | (20) |
| $I_D/I_G$ | <0.9 | 0.9-1 | 1-1.5 | - |
| FWHM-D ($cm^{-1}$) | >170 | 140-170 | <140 | - |

**Table 6**: Metamorphic scale for CV chondrites. Raman parameters defined as in Table 4 and after data of Bonal et al. (2007); NWA 8418 data after MacPherson et al. (2023)

5.5 Other carbonaceous chondrites

While most CK are equilibrated (types 4-6), some of them are of type 3. Although no subtype is officially recognized in the MetBullDB and no scheme has been proposed in the literature, the consensus is that those are >3.5, because of the paucity of refractory inclusions (Greenwood et al. 2010) and the limited heterogeneity of matrix olivine (Dunn et al. 2018; see also Chaumard and Devouard 2016). NWA 5343, which Dunn et al. (2018) deem the least metamorphosed CK, shows a mean chondrule Fa content of 15.9 mol% and a mean matrix olivine Fa content of 36 mol%. If these approximate the (dominant) type I chondrules and type II chondrules respectively, we may estimate a metamorphic grade around M0.5, which would not be too far from Dunn et al. (2018)'s suggested (but uncalibrated) subtype 3.6-3.7. So in our scale too, CK3 would be definitely more metamorphosed than (most) CV3 (Dunn et al. 2018).

---

[14] At face value though, the 0.10±0.12 wt% $Cr_2O_3$ (with 0.28±0.07 wt% MnO, one sigma) reported for Vigarano type II analyses olivine by Grossman and Brearley (2005) (their only extensive dataset for CV chondrites, with 28 analyses), would suggest M0.2 but actually the ratio $Cr_2O_3$/MnO is 0.34±0.07 within error of our general M0.1/M0.2 threshold. The scarce data for other CVs do not allow us to ascertain the initial $Cr_2O_3$ content (with Kaba likely close to the M0.0/M0.1 boundary).

Acfer 094 (C3-an) has $Cr_2O_3/MnO = 1.0$ (Grossman and Brearley 2005) consistent with its very pristine character and a M0.0 grade. For CI chondrites, the 10 ferroan olivine analyses in the dataset of Endress (1994) also yield this value (in agreement with Fig. 3 of Frank et al. 2014) and a M0.0 classification. Likewise, all CM would also be M0.0 according to the data of Schrader and Davidson (2017) and Hewins et al. (2014) plotted on Fig. 8.

Similarly, all CRs are M0.0, in that the type II olivine $Cr_2O_3$ never goes below two thirds of the largest value measured by Schrader et al. (2015), i.e. 0.42±0.07 wt% for EET 96259 (within error of the 0.38 wt% shown by the CR3 chondrites MET 00426 and QUE 99177). GRA 06100 and Y-793495 are marginally below the indicative M0.0/M0.1 threshold $Cr_2O_3/MnO$ value of 0.8 (at 0.73 for both) but this ratio may have been somewhat lower pre-accretion in CRs than for other chondrites (since it exceeds unity only for the anomalous member Al Rais, at 1.1). We nonetheless note that GRA 06100 has undergone some short duration thermal metamorphism (Briani et al. 2013) so the relatively low Cr of its olivine might have significance.

If we combine 13 ferroan olivine analyses of CY chondrite Belgica-7904 from Tomeoka (1990), Kimura and Ikeda (1992), Skirius et al. (1986), we obtain: 0.45±0.20 wt% $Cr_2O_3$ and 0.35±0.15 wt% MnO, hence a $Cr_2O_3/MnO = 1.3±0.2$. This is consistent with the undisturbed type II olivine-chromite igneous partitioning reported by Johnson and Prinz (1991), and the $Fa_I/Fa_{II}=0.03$ calculable from the histograms of Bischoff and Metzler (1991). So CY would be M0.0 too, despite the brief heating event which led to their dehydration (especially for Belgica-7904). Evidently, whether in CY or CR, olivine cannot be held a sensitive recorder of short-lived thermal events, with low integrated diffusion length (although appreciable at submicron scale near crystal boundaries; Nakato et al. 2008). As such, our metamorphic scale probes *protracted* heating (e.g. that due to $^{26}$Al decay), independently of such brief heating events. Thus a separate thermal classification bearing on the brief heating events must be used for them. The Kimura et al. (2011) opaque-based classification in categories A (unheated), B (exsolution of pentlandite), C (exsolution of Ni-rich metal and Co diffusion into kamacite) may be preferable to the more numerous stages I-IV (+ "unheated") of Nakamura (2005) also because they are applicable independently of the preheating hydration state (so e.g. applicable to CR chondrites). Yet, for CY, the Nakamura (2005) XRD criteria should be of course added to the opaque petrographical ones with which they are correlated (A=unheated-stage I; B = II-III; C = III-IV; Kimura et al. 2011), possibly along with Raman properties of organic matter (Briani et al. 2013). So one could classify e.g. Belgica 7904 as M0.0, A0.7TC (or "A0.7TIV" if one sticks to Nakamura (2005)'s heating stages). We note that the opaque (and Raman) criteria for categories B-C may also be fulfilled by protracted metamorphism to 3.1 and beyond (Kimura et al. 2008, 2024), so that by themselves they do not distinguish between brief and protracted heating. This is not an objection to their use for petrologic subtype determination in other chondrite groups (Kimura et al. 2024), since such a disturbance is uncommon, so long the possibility remains to overrule those criteria.

Wild 2 ferroan olivine shows MnO contents comparable to CI and CR olivine (around 0.4 wt%) and the $Cr_2O_3$ values cluster around 0.1 wt% (Frank et al. 2014). This indicates a metamorphic grade of M0.2 for these olivines, that is, if they can be considered representative of chondrules (of which they would be but fragments) when metamorphism occurred (probably not in Wild 2 itself; Frank et al. 2014). This does not mandate a time-temperature history similar to e.g. type 3.2 UOC for the grain size may have been smaller, judging from the few microporphyritic chondrules found (Frank et al. 2014). If the purported parent chondrules were already in fragmented form during metamorphism, the applicability of our chondrule-based classification would be questionable (our own measurement procedure excludes monomineralic fragments), and we might then calibrate it instead on similar-size matrix grains (thus *in fine* on the absolute diffusion length). Comparing with matrix olivine in UOCs, Frank et al. (2014) suggested a subtype of 3.05-3.15.

In Ryugu, Mikouchi et al. (2022) reported olivines clustering around $Fa_1$ and some type II olivine compositions spanning $Fa_{10-56}$ (see also Liu et al. 2022; Kawasaki et al. 2022). This also suggests low metamorphism (M0.0).

### 5.5 Reduced chondrites
The assumptions underlying the operational expression $m=Fa_I/Fa_{II}$ fail in reduced chondrites because they lack *bona fide* type II chondrules. We must return to the first expression in terms of the initial type I and whole-rock olivine composition.

In Kakangari, Berlin (2009) did recognize a "type II trend" among chondrules, with alkali-rich mesostases and sometimes type II-PO-like textures, but neither the Fa (3.5 ± 0.9 mol%) nor the Fs (9.9 ± 3.5 mol%) differ appreciably from those of their "type I trend" (3.6 ± 0.6 mol% and 8.7 ± 4.4 mol%, respectively). The (type I or II) chondrule Fa does differ from matrix olivine (1.5 mol%; Brearley 1989) and the chondrules have incompletely equilibrated with it, with their Fs values presumably approximating $Fa_{I,0}$ because of slower Fe-Mg interdiffusion in pyroxene. If we also equate $Fa_{limit}$ with the bulk composition of more equilibrated K chondrites (e.g. NWA 10085, $Fa_{2.97\pm0.16}$)[15], we obtain $m = 0.89$, hence Kakangari would be a K3.9. We note that application of the Sears et al. (1982) formula using the $RSD(Fa)$ (of chondrule olivine) would put Kakangari at subtype 3.5 (e.g. Berlin 2009) and make the distinction of subtypes dependent on order 0.2 mol% precision in the standard deviation of Fa given the low normalizing average Fa; our convention thus circumvents this.

Enstatite chondrites have no type II chondrule altogether, although some olivine analyses reach up to $Fa_{25}$ (Lusby et al. 1987). However, in EH chondrites, average olivine fayalite contents still show a variation, decreasing from $Fa_{1-2}$ to $Fa_{0.5}$, correlated with the dissolution of olivine (McKibbin et al. 2024), so an olivine compositional definition of $m$ would be nominally applicable. The $Fa_{0.5}$ value is reached around 1 vol% olivine, which Zhang et al. (1995) had already proposed a threshold value between type 3 and 4[16], so it could indeed play the role of $Fa_{limit}$. Yet, uncertainties in the evaluation of $Fa_I$ itself from so minor a EH component phase make it an unpractical assignation criterion, only to be formally correlated to more operational ones, e.g. olivine *modal abundance*. It is still uncertain whether any known enstatite chondrite should count as a M0.0. The organic matter maturity of Sahara 97096—the least advanced reported by Quirico et al. (2011)—is intermediate between type 3.0 and 3.1 ordinary chondrites, although, again, the precursor need not have been the same. The low matrix-normalized abundances of presolar silicates in enstatite chondrites (Ebata et al. 2006) likewise leave room for even less metamorphosed samples. Possibly though their major silicate mineralogy and chemistry may not be too different from those observed in Sahara 97096 (~10 vol% olivine; McKibbin et al. 2024). The Fs histograms of Lusby et al. (1987) suggest that pyroxene (less sensitive to diffusion) of UEC has a mean Mg# around 98-99 comparable to the olivine in those meteorites, so olivine was likely not more oxidized or abundant than this upon accretion. So one might define a robust olivine mode-based classification, e.g. something like $m \approx 1-(x_{Ol}/10\ \text{vol}\%)$, with $x_{Ol}$ the olivine fractional abundance, which would be rounded up to 1 for $x_{Ol}$ below 0.5 vol%, i.e. about the type 3/type 4 transition upheld above. Perhaps the olivine/pyroxene ratio could be more robust to fluctuations of the abundance of opaque nodules. Unfortunately, there is little overlap between the samples analyzed by Quirico et al. (2011) and McKibbin et al. (2024), so a metamorphic subclassification of EH3, with different correlated assignation parameters (Raman properties, olivine composition and abundance, presence of ferroan pyroxene, presolar grain abundances…), is not yet at hand. The same *a fortiori* can be said of EL3 chondrites.

---

[15] A better proxy would be the average olivine composition in whole-rock Kakangari itself (matrix+chondrules) but we would need to know the modal abundance of olivine in both matrix (~20 vol% ; Brearley 1989) and chondrules.

[16] The transition cannot be defined with olivine variability (as 5 PMD FeO would be hard to resolve in standard microprobe surveys for such forsteritic compositions), as pointed out from the outset by Van Schmus and Wood (1967).

## 6. Outlook and perspectives

Our metamorphic and aqueous alteration scales produce the two-dimensional classification shown in Fig. 14. When the coordinates are understood as continuous variables, chondrites are never off by more than ~0.05 from either the *x* or the *y* axis. This means that they are always either A0.0 or M0.0. Clearly, this apparent exclusivity of metamorphism and aqueous alteration is a matter of the chosen precision of the taxonomy (and its metrics). The very heating which caused melting of ice and aqueous alteration also caused incipient diffusion and recrystallization in the primary phases such as olivine. However, aqueous alteration in CM chondrites, for example, took place at temperatures variously estimated at 0 to 300°C (Suttle et al. 2021b), which would not allow Fe-Mg diffusion lengths in olivine longer than 0.01 µm over a Ma timescale (Schwinger et al. 2016). Yet we mentioned earlier the possibility of measurable Cr depletion in type II chondrule olivine in some CR chondrites (Fig. 5). As to noncarbonaceous or CV chondrites, the low amount of fine-grained matrix and water limited the amount of phyllosilicates produced but these are present up to M0.2 in OC and CV chondrites. Higher subtypes may have lost phyllosilicates by dehydration.

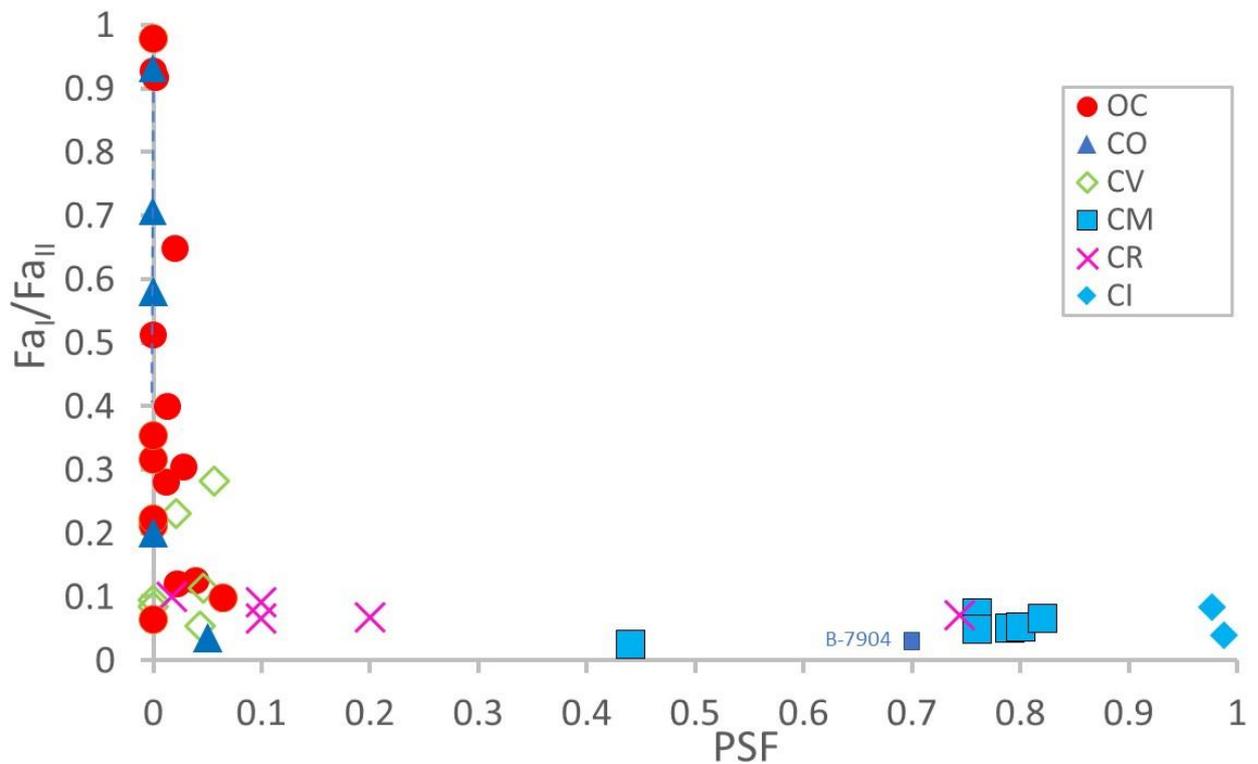

**Figure 14**: Recapitulative biplot of *m* versus the PSF. Data from this study or literature cited in previous figures, with the following additions: Howard et al. (2010) for CV PSF values, King et al. (2015) for CI's, and Grant et al. (2023) for UOC's (we crudely calculated a maximum phyllosilicate abundance by assuming their bulk water content measured to be entirely in the form of phyllosilicates with 20 wt% water). A PSF value of zero has been assigned to CO (save for the exceptions with detectable phyllosilicates in Alexander et al. 2018) and Acfer 187/El Djouf 001 and Renazzo are assigned rounded PSF values of 0.1 and 0.2, respectively, following Table 3. Additional Fa data are from Endress (1994) for CI chondrites, Bischoff et al. (1993) for Acfer 187 and El Djouf 001, Hewins et al. (2014) for Paris, Hutchison and Symes (1972) for Murchison, Wood (1967) for other CM and CR chondrites.

This apparent exclusivity of (taxonomically notable) aqueous and metamorphic alteration leaves the taxonomy *de facto* univariant (notwithstanding a possible brief heating descriptor as discussed for CY chondrites). Hence, a single Van Schmus and Wood (1967) petrologic subtype can still convey the essential information as in present MetBullDB usage. For type 3 chondrites (A0.0), the subtype can be 3+r($m$) (with, again, "r" meaning rounding to the relevant tenth), without ambiguity. Indeed, on the whole, our redefinition is a minor correction to existing schemes, especially for ordinary chondrites (Sears et al. 1980; Grossman and Brearley 2005).

As to type 1 and 2 chondrites, it would be tempting to follow Howard et al. (2015) and define petrologic subtypes as 3-2*PSF* rounded to the nearest tenth, as a transition between 3 and 1. However, unless one artificially restricts subtypes to even tenths (3.0, 2.8, 2.6, 2.4 etc.), this would motivate an overprecise, and thus less robust classification (equivalent to distinguishing e.g. A0.75, A0.80, A0.85…) which we have argued against in the section The aqueous alteration scale. We prefer to compress the scale to the [2.0;3] interval, similar to Rubin et al. (2007) and Harju et al. (2014), and define petrologic subtypes as 3-r(*PSF*). So type 1 would encompass subtypes 2.1 and 2.0. It must be cautioned though that while we adopt the [2.0;3] interval of Rubin et al. (2007) and Harju et al. (2014), our PSF-defined subtypes coincide with theirs only near types 1 and 3 (Tables 2-3), even though the underlying assignation criteria embody their work. So one should be mindful of possible confusions.

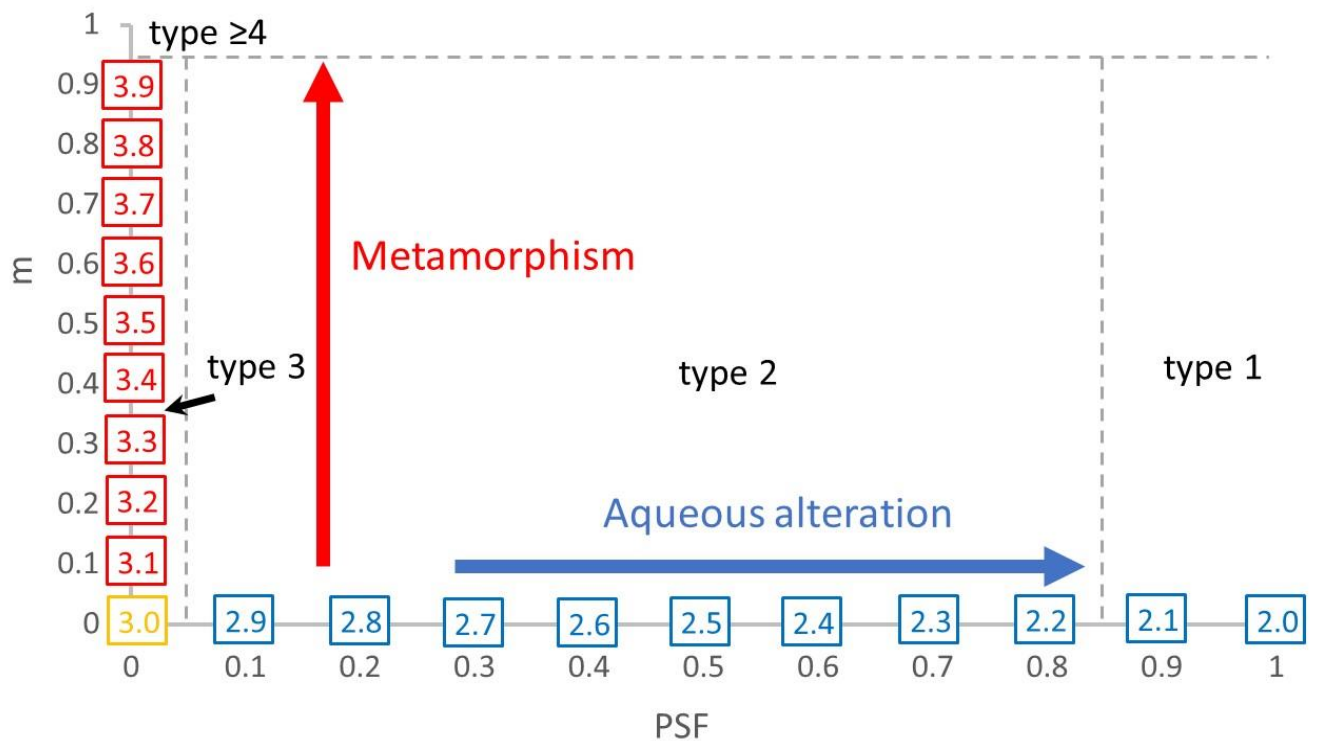

**Figure 15**: Schematic plot of the conversion of our bidimensional subclassification in Van Schmus-Wood subtypes. At the level of precision chosen, the chondrites indeed form a one-dimensional line that can be essentially parameterized by one decimal subtype. Dashed lines show boundaries of Van Schmus and Wood (1967) *integer* petrologic types 1-4.

In summary then, the general petrologic subtype would be 3+r(*m*)-r(*PSF*). This is depicted in Fig. 15. This does not prejudge the possibility of more subdivisions in the alteration or metamorphic scales, e.g. A0.0 in A0.00 and A0.05; or M0.0 in M0.00 and M0.05 (e.g. the type 3.00 and 3.05 of Grossman and Brearley (2005) or Rubin and Li (2019)). These would no longer be capturable in the single petrologic subtype above, which warrants the formal development of the separate scales. The extra information could be added parenthetically, e.g. the Bali-type oxidized CV chondrite Grosnaja would be a "$CV_{ox}3.2$ (A0.05)".

We have already commented for both scales that genomict breccias can be indicated with hyphens between the represented metamorphic and aqueous alteration grades (so long they account for >10 vol%) as well as the Van Schmus and Wood (1967) types. For both scales, a bulk value roughly means a volume-weighted average of the values of the individual lithologies. So this bulk value (with possibly the range) makes sense even for them. Thus our adopted alteration and metamorphic degrees play a fully symmetric role.

Our purpose has been to provide a unified, yet flexible framework for the secondary classification of chondrites, molded after existing practice grafted on the Van Schmus and Wood (1967) scale. Precise boundaries between alteration or metamorphic grades and the relevant parameters must be adapted to chondrite groups of interest. We have proposed such adaptations for several groups mostly based on existing literature; in fact our classifications may be merely seen as modifications of existing ones recast in our new framework. Other such classifications can be devised for other groups (e.g. R, E, CK chondrites…) by dedicated studies and none of the existing ones should be of course exempt from revision. Yet if the present framework is adopted, assigned grades would be robust and unlikely to change by more than 0.1.

In this purely taxonomic work, we have not attempted to gain new insights on secondary processes. The metrics chosen (olivine equilibration, abundance of nominally hydrous silicates) are the most obvious and safest ones possible, whatever the exact context of the processes responsible. Yet we hope that our proposed rationalization of secondary taxonomy will prove a practical tool to better unravel the postaccretionary history of chondrite parent bodies.

## Supporting information:

EA_olivine: Individual olivine data from this study.

Supplements A-E: Supplementary text.

*Acknowledgments*: We are grateful to Kevin Righter, Makoto Kimura, and an anonymous reviewer for their constructive comments. We thank Jeffrey Grossman, Roger Hewins, Rhian Jones, Tim McCoy, Derek Sears, Martin Suttle for providing access to literature data. This paper is dedicated to the memory of the lead author's *beau-grand-père* Georges Illy.

## References

Abreu N. M. (2016). Why is it so difficult to classify Renazzo-type (CR) carbonaceous chondrites? - Implications from TEM observations of matrices for the sequences of aqueous alteration. *Geochimica et Cosmochimica Acta* 194:91-122.

Abreu N. M. Brearley A. J. (2010). Early solar system processes recorded in the matrices of two highly pristine CR3 carbonaceous chondrites, MET 00426 and QUE 99177. *Geochimica et Cosmochimica Acta* 74:1146-1171.

Abreu N. M., Aponte J. C., Nguyen A. N. (2020). The Renazzo-like carbonaceous chondrites as resources to understand the origin, evolution, and exploration of the solar system. *Geochemistry* 80:125631.

Alexander C. M. O'D., Ebel D. S. (2012). Questions, questions: Can the contradictions between the petrologic, isotopic, thermodynamic, and astrophysical constraints on chondrule formation be resolved? *Meteoritics and Planetary Science* 47:1157-1175.

Alexander C. M. O'D., Howard K. T., Bowden R., Fogel M. L. (2013). The classification of CM and CR chondrites using bulk H, C, and N abundances and isotopic compositions. *Geochimica et Cosmochimica Acta* 123:244-260.

Alexander C. M. O'D., Greenwood R. C., Bowden R., Gibson J. M., Howard K. T., Franchi I. A. (2018). A multi-technique search for the most primitive CO chondrites. *Geochimica et Cosmochimica Acta* 221:406-420.

Bendersky C., Weisberg M. K., Connolly H. C., Ebel D. S. (2007). Olivine and the Onset of Thermal Metamorphism in EH3 Chondrites. XXXVIII[th] *Lunar and Planetary Science Conference* abstract #2077. CD-ROM.

Berlin J. (2009). Mineralogy and bulk chemistry of chondrules and matrix in petrologic type 3 chondrites: Implications for early solar system processes. PhD thesis, Freie Universität Berlin.

Binns R. A. (1968). Cognate xenoliths in chondritic meteorites: Examples in Mezö-Madaras and Ghubara. *Geochimica et Cosmochimica Acta* 32:299-317.

Bischoff A., Metzler K. (1991). Mineralogy and petrography of the anomalous carbonaceous chondrites Yamato-86720, Yamato-82162, and Belgica-7904. *Proceedings of the NIPR Symposium on Antarctic Meteorites* 15:226-245.

Bischoff A. (2000). Mineralogical characterization of primitive, type-3 lithologies in Rumuruti chondrites. *Meteoritics and Planetary Science* 35:699-706.

Bischoff A., Palme H., Ash R. D., Clayton R. N., Schultz L., Herpers U., Stöffler D., Grady M. M., Pillinger C. T., Spettel B., Weber H., Grund T., Endress M., Weber D. (1993). Paired Renazzo-type (CR) carbonaceous chondrites from the Sahara. *Geochimica et Cosmochimica Acta* 57:1587-1603.

Bischoff A., Schleiting M., Wieler R., Patzek M. (2018). Brecciation among 2280 ordinary chondrites—Constraints on the evolution of their parent bodies. *Geochimica et Cosmochimica Acta* 238:516-541.

Bischoff A., Schleiting M., Patzek M. (2019). Shock stage distribution of 2280 ordinary chondrites—Can bulk chondrites of shock stage of S6 exist as individual rocks? *Meteoritics and Planetary Science* 54:2189-2202.

Bonal L., Bourot-Denise M., Quirico E., Montagnac G., Lewin E. (2006). Determination of the petrologic type of CV3 chondrites by Raman spectroscopy of included organic matter. *Geochimica et Cosmochimica Acta* 70:1849-1863.

Bonal L., Bourot-Denise M., Quirico E., Montagnac G., Lewin E. (2007). Organic matter and metamorphic history of CO chondrites. *Geochimica et Cosmochimica Acta* 71:1605-1623.

Bonal L., Quirico E., Flandinet L., Montagnac G. (2016). Thermal history of type 3 chondrites from the Antarctic meteorite collection determined by Raman spectroscopy of their polyaromatic carbonaceous matter. *Geochimica et Cosmochimica Acta* 189:312-337.

Brearley A. J. (1996). Grain Size Distribution and Textures in the Matrices of Metamorphosed CO Chondrites. *Lunar and Planetary Science Conference* 27:159-160.

Brearley A.J. (2014). Nebular Versus Parent Body Processing. In Treatise on Geochemistry, editors K. Turekian, H. Holland, chapter 1.9, Elsevier, Paris.

Brearley A. J., Jones R. H. (1998). Chondritic Meteorites. In Planetary Materials, editor J. J. Papike, Reviews of Mineralogy and Geochemistry, 3-1-3-398, Mineralogical Society of America, Washington D. C.

Briani G., Quirico E., Gounelle M., Paulhiac-Pison M., Montagnac G., Beck P., Orthous-Daunay F.-R., Bonal L., Jacquet E., Kearsley A., Russell S. S. (2013). Short duration thermal metamorphism in CR chondrites. *Geochimica et Cosmochimica Acta* 122:267-279.

Browning L. B., McSween H. Y. Jr., Zolensky M. E. (1996). Correlated alteration effects in CM carbonaceous chondrites. *Geochimica et Cosmochimica Acta* 60:2621-2633.

Bunch T. E., Keil K., Snetsinger K. G. (1967). Chromite composition in relation to chemistry and texture of ordinary chondrites. *Geochimica et Cosmochimica Acta* 31:1569-1582.

Chakraborty S. (2010). Diffusion Coefficients in Olivine, Wadsleyite and Ringwoodite. In Diffusion in Minerals and Melts, editors Zhang Y. and Cherniak D. J., Reviews in Mineralogy and Geochemistry, Mineralogical Society of America, Washington D. C., 603-639.

Chaumard N., Devouard B. (2016). Chondrules in CK carbonaceous chondrites and thermal history of the CV-CK parent body. *Meteoritics and Planetary Science* 51:547-573.

Chizmadia L., Rubin A. E., Wasson J. T. (2002). Mineralogy and petrology of amoeboid olivine inclusions in CO3 chondrites: Relationship to parent body aqueous alteration. *Meteoritics and Planetary Science* 37:1781-1796.

Christophe Michel-Levy M. (1969). Etude minéralogique de la chondrite CIII de Lancé. In: Millman, P.M. (eds) Meteorite Research. Astrophysics and Space Science Library, vol 12. Springer, Dordrecht.

Christophe Michel-Levy M. (1976). La matrice noire et blanche de la chondrite de Tieschitz (H3). *Earth and Planetary Science Letters* 30:143-150.

Ciocco M., Roskosz M., Doisneau B., Beyssac O., Mostefaoui S., Remusat L., Leroux H., Gounelle M. (2022). Impact dynamics of the L chondrites' parent asteroid. *Meteoritics and Planetary Science* 57:759-775.

Crank J. (1975). The Mathematics of Diffusion. Oxford University Press, Oxford.

Davidson J., Alexander C. M. O'D., Stroud R. M., Busemann H., Nittler L. R. (2019). Mineralogy and petrology of Dominion Range 08006: A very primitive CO3 carbonaceous chondrite. *Geochimica et Cosmochimica Acta* 265:259-278.

DeHart J. M. (1989). Cathodoluminescence and microprobe studies of the unequilibrated ordinary chondrites. PhD thesis, University of Arkansas.

DeHart J. M., Lofgren G. E., Jie L., Benoit P. H., Sears D. W. G. (1992). Chemical and physical studies of chondrites: X. Cathodoluminescence and phase composition studies of metamorphism and nebular processes in chondrules of type 3 ordinary chondrites. *Geochimica et Cosmochimica Acta* 56:3791-3807.

Dobrica E., Brearley A. J. (2014). Widespread hydrothermal alteration minerals in the fine-grained matrices of the Tieschitz unequilibrated ordinary chondrite. *Meteoritics and Planetary Science* 49:1323-1349.

Dodd R. T., Jr., Koffman D. M., Van Schmus W. R. (1967). A survey of the unequilibrated ordinary chondrites. *Geochimica et Cosmochimica Acta* 31:921-951.

Dundon R. W., Walter L. S. (1967). Ferrous ion order-disorder in meteoritic pyroxenes and the metamorphic history of chondrites. *Earth and Planetary Science Letters* 2:372-376.

Dunn T. L., Cressey G., McSween H. Y. Jr., McCoy T. J. (2010). Analysis of ordinary chondrites using powder X-ray diffraction: 1. Modal mineral abundances. *Meteoritics and Planetary Science* 45:123-134.

Dunn T. L., Battifarano O. K., Gross J., O'Hara E. (2018). Characterization of matrix material in Northwest Africa 5343: Weathering and thermal metamorphism of the least equilibrated CK chondrite. *Meteoritics and Planetary Science* 53:2165-2180.

Ebata S., Nagashima K., Itoh S., Kobayashi S., Sakamoto N., Fagan T. J., Yurimoto H. (2006). Presolar silicate grains in enstatite chondrites. *Lunar and Planetary Science Conference XXXVII*, abstract #1619. CD-ROM.

Frank D. R., Zolensky M. E., Le L. (2014). Olivine in terminal particles of Stardust aerogel tracks and analogous grains in chondrite matrix. *Geochimica et Cosmochimica Acta* 142:240-259.

Gattacceca J., Bonal L., Sonzogni C., Longerey J. (2020). CV chondrites: More than one parent body. *Earth and Planetary Science Letters* 547:116467.

Gordon S. H. (2009). The composition of primitive meteorites. PhD thesis. Imperial College London.

Grant H., Tartèse R., Jones R., Piani L., Marrocchi Y., King A., Rigaudier T. (2023). Bulk mineralogy, water abundance, and hydrogen isotope composition of unequilibrated ordinary chondrites. *Meteoritics and Planetary Science* 58:1365-1381.

Greenwood R. C., Franchi I. A., Kearsley A. T., Alard O. (2010). The relationship between CK and CV chondrites. *Geochimica et Cosmochimica Acta* 74:1684-1705.

Gross J., Treiman A. H., Connolly H. C. Jr. (2017). Water on asteroids: The curious case of R-chondrite Muller Range 11207. 80[th] Annual Meeting of The Meteoritical Society. Abstract #1987. CD-ROM.

Grossman J. N., Brearley A. J. (2005). The onset of metamorphism in ordinary and carbonaceous chondrites. *Meteoritics and Planetary Science* 40:87-122.

Grossman J. N., Rubin A. E., Sears D. W. G. (2009). The Mineral Compositions and Classification of High Type-3 and Type-4 Ordinary Chondrites. *40th Annual Lunar and Planetary Science Conference*. Abstract #1679. CD-ROM.

Grossman J. N. (2011). Classification of Ordinary Chondrites Based on Mean and Standard Deviation of Fa and Fs contents of Mafic Silicates. https://www.lpi.usra.edu/meteor/docs/whitepaper-supp.pdf.

Guimon R. K., Symes S. J. K., Sears D. W. G., Benoit P. H. (1995). Chemical and physical studies of type 3 chondrites XII: The metamorphic history of CV chondrites and their components. *Meteoritics* 30:704-714.

Haenecour P., Floss C., Zega T. J., Croat T. K., Wang A., Jolliff B. L., Carpenter P. (2018). Presolar silicates in the matrix and fine-grained rims around chondrules in primitive CO3.0 chondrites: Evidence for pre-accretionary aqueous alteration of the rims in the solar nebula. *Geochimica et Cosmochimica Acta* 221:379-405.

Han J., Park C., Brearley A. J. (2022). A record of low-temperature asteroidal processes of amoeboid olivine aggregates from the Kainsaz CO3.2 chondrite. *Geochimica et Cosmochimica Acta* 322:109-128.

Hanowski N. P., Brearley A. J. (2001). Aqueous alteration of chondrules in the CM carbonaceous chondrite, Allan Hills 81002: implications for parent body alteration. *Geochimica et Cosmochimica Acta* 65:495-518.

Harju E. R., Rubin A. E., Ahn I., Choi B.-G., Ziegler K., Wasson J. T. (2014). Progressive aqueous alteration of CR carbonaceous chondrites. *Geochimica et Cosmochimica Acta* 139:267-292.

Hewins R. H., Bourot-Denise M., Zanda B., Leroux H., Barrat J.-A., Humayun M., Göpel C., Greenwood R. C., Franchi I. A., Pont S., Lorand J.-P., Cournède C., Gattacceca J., Rochette P., Kuga M., Marrocchi Y., Marty B. (2014). The Paris meteorite, the least altered CM chondrite so far. *Geochimica et Cosmochimica Acta* 124:190-222.

Howard K. T., Benedix G. K., Bland P. A., Cressey G. (2010). Modal mineralogy of CV3 chondrites by X-ray diffraction (PSD-XRD). *Geochimica et Cosmochimica Acta* 74:5084-5097.

Howard K. T., Alexander C. M. O'D., Dyl K. A. (2014). PSD-XRD Modal Mineralogy of Type 3.0 CO Chondrites: Initial Asteroidal Water Mass Fractions and Implications for CM Chondrites. 45th Annual Lunar and Planetary Science Conference. Abstract #1830. CD-ROM.

Howard K. T., Alexander C. M. O'D., Schrader D. L., Dyl K. A. (2015). Classification of hydrous meteorites (CR, CM and C2 ungrouped) by phyllosilicate fraction: PSD-XRD modal mineralogy and planetesimal environments. *Geochimica et Cosmochimica Acta* 149:206-222.

Howard K. T., Zanda B. (2019). Why is the Degree of Aqueous Alteration Variable? 82$^{nd}$ Annual Meeting of The Meteoritical Society. Abstract #6178. CD-ROM.

Huss G. R., Rubin A. E., Grossman J. N. (2006). Thermal Metamorphism in Chondrites. In *Meteorites and the Early Solar System II*, p 567-586, eds. D. S. Lauretta and H. Y. McSween Jr, Cambridge University Press.

Hutchison R., Symes R. F. (1972). Calcium Variation in Olivines of the Murchison and the Vigarano Meteorites. *Meteoritics* 7:23-47.

Hutchison R., Alexander C.M.O'D., Barber D.J. (1987). The Semarkona meteorite: First recorded occurrence of smectite in an ordinary chondrite, and its implications. *Geochimica et Cosmochimica Acta* 51:1875-1882.

Ikeda Y. (1992) An overview of the research consortium, antarctic carbonaceous chondrites with CI affinities, Yamato-86720, Yamato-82162, and Belgica-7904. *Proceedings of the NIPR Symposium on Antarctic Meteorites* 5 :49-73.

Jacquet E., Alard O., Gounelle M. (2012). Chondrule trace element geochemistry at the mineral scale. *Meteoritics and Planetary Science* 47 :1695-1714.

Jacquet E., Alard O., Gounelle M. (2015). Trace element geochemistry of ordinary chondrite chondrules: The type I/type II chondrule dichotomy. *Geochimica et Cosmochimica Acta* 155:47-67.

Jacquet E., Barrat J.-A., Beck P., Caste F., Gattacceca J., Sonzogni C., Gounelle M. (2016). Northwest Africa 5958: a weakly altered CM-related ungrouped chondrite, not a CI3. *Meteoritics and Planetary Science* 51:851-869.

Jacquet E., Piralla M., Kersaho P., Marrocchi Y. (2021). Origin of isolated olivine grains in carbonaceous chondrites. *Meteoritics and Planetary Science* 56:13-33.

Jacquet E. (2022). Meteorite petrology versus genetics: Toward a unified binominal classification. *Meteoritics and Planetary Science* 57:1774-1794.

Jarosewich E. (1990). Chemical analyses of meteorites: A compilation of stony and iron meteorite analyses. *Meteoritics* 25:323-337.

Jing J.-J., Su B.-X., Berndt J., Kuwahara H., van Westrenen W. (2024). Experimental investigation of first-row transition elements partitioning between olivine and silicate melt: Implications for lunar basalt formation. *Geochimica et Cosmochimica Acta* 373:211-231.

Johnson C. A., Prinz M. (1991). Chromite and olivine in type II chondrules in carbonaceous and ordinary chondrites: Implications for thermal histories and group differences. *Geochimica et Cosmochimica Acta* 55:893-904.

Jones R. H. (1992). On the relationship between isolated and chondrule olivine grains in the carbonaceous chondrite ALHA77307. *Geochimica et Cosmochimica Acta* 56:467-482.

Jones R. H. (1994). Petrology of FeO-poor, porphyritic pyroxene chondrules in the Semarkona chondrite. *Geochimica et Cosmochimica Acta* 58:5325-5340.

Kallemeyn G. W., Rubin A. E., Wasson J. T. (1991). The compositional classification of chondrites: V. The Karoonda (CK) group of carbonaceous chondrites. *Geochimica et Cosmochimica Acta* 55:881-892.

Kawasaki, N. ; Nagashima, K. ; Sakamoto, N. ; Matsumoto, T. ; Bajo, K. ; Wada, S. ; Igami, Y.  Miyake, Ak. ; Noguchi, Takaaki ; Yamamoto, D.; Russell, S. S. ; Abe, Y. ; Aléon, J. ;  Alexander, C. M. O. 'D. ; Amari, S. ; Amelin, Y. ; Bizzarro, M. ; Bouvier, A. ; Carlson, R. W. ; Chaussidon, M. , Choi, B.-G. ; Dauphas, N. ; Davis, A. M. ; Di Rocco, T. ; Fujiya, W. ; Fukai, R. ; Gautam, I. ; Haba, M. K. ; Hibiya, Y. ; Hidaka, H. ; Homma, H. ; Hoppe, P. ; Huss, G. R. ; Ichida, K. ; Iizuka, T. ; Ireland, T. R. ; Ishikawa, A. ; Ito, M. ; Itoh, S. ; Kita, N. T. ; Kitajima, K. ; Kleine, T. ; Komatani, S. ; Krot, A. N. ; Liu, M.-C. ; Masuda, Y. ; McKeegan, K. D. ; Morita, M. ; Motomura, K. ; Moynier, F. ; Nakai, I. ; Nguyen, A. ; Nittler, L. ; Onose, M. ; Pack, A. ; Park, C. ; Piani, L. ; Qin, L. ; Schönbächler, M. ; Tafla, L. ; Tang, H. ; Terada, K. ; Terada, Y. ; Usui, T. ; Wadhwa, M. ; Walker, R. J. ; Yamashita, K. ; Yin, Q.-Z. ; Yokoyama, T. ; Yoneda, S. ; Young, E. D. ; Yui, H. ; Zhang, A.-C. ; Nakamura, T. ; Naraoka, H.

; Okazaki, R. ; Sakamoto, K. ; Yabuta, H. ; Abe, M. ; Miyazaki, A. ; Nakato, A. ; Nishimura, M. ; Okada, T. ; Yada, T. ; Yogata, K. ; Nakazawa, S. ; Saiki, T. ; Tanaka, S. ; Terui, F. ; Tsuda, Y. ; Watanabe, S. ; Yoshikawa, M. ; Tachibana, S. ; Yurimoto, H. (2022). Oxygen isotopes of anhydrous primary minerals show kinship between asteroid Ryugu and comet 81P/Wild 2. Science Advances 8 :eade2067.

Keil K. (1964). Mineralogical and Chemical Relationships among Enstatite Chondrites. *Journal of Geophysical Research* 73:6945-6976.

Keil K., Mason B., Wiik H. B., Frederiksson K. (1964). The Chainpur meteorite. *American Museum Novitate* 2173.

Kimura M., Ikeda Y. (1992). Mineralogy and Petrology of an unusual Belgica-7904 carbonaceous chondrite: genetic relationships among the components. *Proceedings of the NIPR Symposium on Antarctic Meteorites* 5:74-119.

Kimura M., Hiyagon H., Palme H., Spettel B., Wolf D., Clayton R. N., Mayeda T. K., Sato T., Suzuki A., Kojima H. (2002). Yamato 792947, 793408 and 82038: The most primitive H chondrites, with abundant refractory inclusions. *Meteoritics and Planetary Science* 37:1417-1434.

Kimura M., Grossman J.N. and Weisberg M.K. (2008) Fe-Ni metal in primitive chondrites: Indicators of classification and metamorphic conditions for ordinary and CO chondrites. *Meteoritics and Planetary Science* 43:1161-1177.

Kimura M., Grossman J. N., Weisberg M. K. (2011). Fe-Ni metal and sulfide minerals in CM chondrites: An indicator for thermal history. *Meteoritics and Planetary Science* 46:431-442.

Kimura M., Imae N., Komatsu M., Barrat J.A., Greenwood R.C., Yamaguchi A., Noguchi T. (2020). The most primitive CM chondrites, Asuka 12085, 12169, and 12236, of subtypes 3.0-2.8: Their characteristic features and classification. *Polar Science* 26:100565.

Kimura M. et al. (2022). Petrology and classification of A‐9003, A 09535, and Y‐82094: A new type of carbonaceous chondrite. *Meteoritics and Planetary Science* 57:302-316.

Kimura M., Weisberg M. K., Yamaguchi A. (2024). Subtype 3.0 chondrites: Petrologic classification criteria. *Meteoritics and Planetary Science* 59:858-877.

King A. J., Schofield P.F., Howard K. T., Russell S. S. (2015). Modal mineralogy of CI and CI-like chondrites by X-ray diffraction. *Geochimica et Cosmochimica Acta* 165:148-160.

King A. J., Bates H. C., Krietsch D., Busemann H., Clay P.L., Schifield P.F., Russell S.S. (2019). The Yamato-type (CY) carbonaceous chondrite group: Analogues for the surface of asteroid Ryugu? *Geochemistry* 79:125531.

King A.J., Schofield P.F. and Russell S.S. (2021) Thermal alteration of CM carbonaceous chondrites: Mineralogical changes and metamorphic temperatures. *Geochimica et Cosmochimica Acta* 298 :167-190.

Kracher A., Keil K., Kallemeyn G. W., Wasson J. T., Clayton R. N., Huss G. I. (1985). The Leoville (CV3) Accretionary Breccia. *Journal of Geophysical Research* 90:D123-D135.

Krämer-Ruggiu L., Devouard B., Gattacceca J., Bonal L., Leroux H., Eschrig J., Borschneck D., King A. J., Beck P., Marrocchi Y., Debaille V., Hanna R. D., Grauby O. (2022). Detection of incipient aqueous alteration in carbonaceous chondrites. *Geochimica et Cosmochimica Acta* 336:308-331.

Krot A. N., Keil K., Scott E. R. D. (2014). Classification of Meteorites and Their Genetic Relationships. In *Treatise on Geochemistry*, vol. 1, p 1-53, eds. K. Turekian and H. Holland, Elsevier.

Le Guillou C., Changela H. G., Brearley A. J. (2015). Widespread oxidized and hydrated amorphous silicates in CR chondrites matrices: Implications for alteration conditions and $H_2$ degassing of asteroids. *Earth and Planetary Science Letters* 420:162-173.

Leitner J., Hoppe P., Zipfel J. (2015). Distribution and Abundance of Presolar Silicate and Oxide Stardust in CR Chondrites. *46$^{th}$ Lunar and Planetary Science Conference*, held March 16-20, 2015 in The Woodlands, Texas. LPI Contribution No. 1832, p.1874.

Lentfort S., Bischoff A., Ebert S., Patzek M. (2021). Classification of CM chondrite breccias—Implications for the evaluation of samples from the OSIRIS-Rex and Hayabusa 2 missions. *Meteoritics and Planetary Science* 56:127-147.

Li H. Y. (2000). A compendium of geochemistry—From solar nebula to the human brain. Princeton, New Jersey, Princeton University Press.

Liu M.-C., McCain K. A., Matsuda N., Yamaguchi A., Kimura M., Tomioka N., Ito M., Uesugi M., Imae N., Shirai N., Ohigashi T., Greenwood R. C., Uesugi K., Nakato A., Yogata H., Kodama Y., Hirahara K., Sakurai I., Okada I., Karouji Y., Nakazawa S., Okada T., Saiki T., Tanaka S., Terui F., Yoshikawa M., Miyazaki A., Nishimura M., Yada T., Abe Masanao, Usui T., Watanabe S., Tsuda Y. (2022). Incorporation of $^{16}$O-rich anhydrous silicates in the protoliths of highly hydrated asteroid Ryugu. *Nature Astronomy* 6:1172-1177.

Lodders K. (2003). Solar System Abundances and Condensation Temperatures of the Elements. *The Astrophysical Journal* 591:1220-1247.

Lusby D., Scott E. R. D., Keil K. (1987). Ubiquitous High-FeO Silicates in Enstatite Chondrites. Proceedings of the Seventeenth Lunar and Planetary Science Coference, Part 2, *Journal of Geophysical Research*, 92:E679-E695.

Ma N., Neumann W., Néri A., Schwarz W. H., Ludwig T., Trieloff M., Klahr H., Bouvier A. (2022). Early formation of primitive achondrites in an outer region of the protoplanetary disc. *Geochemical Perspective Letters* 23:33-37.

MacPherson G.J., Nagashima K., Krot A. N., Kuehner S.M., Irving A. J., Ziegler K., Mallozzi L., Corrigan C., Pitt D. (2023). Northwest Africa 8418: The first CV4 chondrite. *Meteoritics and Planetary Science* 58:135-157.

Marrocchi Y., Gounelle M., Blanchard I., Caste F., Kearsley A. T. (2014). The Paris CM chondrite: Secondary minerals and asteroidal processing. *Meteoritics and Planetary Science* 49:1232-1249.

Marrocchi Y., Piralla M., Regnault M., Batanova V., Villeneuve J., Jacquet E. (2022). Isotopic evidence for two chondrule generations in CR chondrites and their relationships to other carbonaceous chondrites. *Earth and Planetary Science Letters* 593:117683.

Marrocchi Y., Jacquet E., Neukampf J., Villeneuve J., Zolensky M. E. (2023). From whom Bells tolls: Reclassifying Bells among CR chondrites and implications for the formation conditions of CR parent bodies. *Meteoritics and Planetary Science* 58 :195-206.

Martinez M., Brearley A. J. (2022). Smooth rims in Queen Alexandra Range (QUE) 99177: Fluid-chondrule interactions and clues on the geochemical conditions of the primordial fluid that altered CR carbonaceous chondrites. *Geochimica et Cosmochimica Acta* 325:39-64.

Mason B. (1962). *Meteorites*. Wiley, New York.

Matsunami S., Nishimura H., Takeshi H. (1990). The chemical compositions and textures of matrices and chondrule rims of unequilibrated ordinary chondrites-II Their constituents and the implications for the formation of matrix olivine. *Proceedings of the NIPR Symposium on Antarctic Meteorites* 3:147-180.

McCanta M. C., Treiman A. H., Dyar M. D., Alexander C.M.O'D., Rumble D. III, Essene E.J. (2008). The LaPaz Icefield 04580 meteorite: Mineralogy, metamorphism and origin of an amphibole- and biotite-bearing R chondrite. *Geochimica et Cosmochimica Acta* 72:5757-5780.

McCoy T. J., Scott E. R. D., Jones R. H., Keil K., Taylor G. J. (1991). Composition of chondrule silicates in LL3-5 chondrites and implications for their nebular history and parent body metamorphism. *Geochimica et Cosmochimica Acta* 55:601-619.

McKibbin S. J., Hecht L., Makarona C., Huber M., Terryn H., Claeys P. (2024). Forsteritic olivine in EH (enstatite) chondrite meteorites: A record of nebular, metamorphic, and crystal-lattice diffusion effects. *Meteoritics and Planetary Science*.

McSween H. Y., Jr. (1977a). Carbonaceous chondrites of the Ornans type: A metamorphic sequence. *Geochimica et Cosmochimica Acta* 41:477-491.

McSween H. Y., Jr. (1977b). Petrographic variations among carbonaceous chondrites of the Vigarano type. *Geochimica et Cosmochimica Acta* 41:1777-1790.

McSween H. Y., Jr. (1979). Alteration in CM carbonaceous chondrites inferred from modal and chemical variations in matrix. *Geochimica et Cosmochimica Acta* 43:1761-1770.

Metzler K., Hezel D. C., Barosch J., Wölfer E., Schneider J. M., Helmann J. L., Berndt J., Stracke A., Gattacceca J., Greenwood R. C., Franchi I. A., Burkhardt C., Kleine T. (2021). The Loongana (CL) group of carbonaceous chondrites. *Geochimica et Cosmochimica Acta*, 304:1-31.

Mikouchi T., Nakamura T.; Zolensky M. E., Yoshida H., Najashima D., Hagiya K., Kikuiri M., Morita T., Amano K., Kagawa E., Yurimoto H., Noguchi T., Okazaki R., Yabuta H., Naraoka H., Sakamoto K., Tachibana S., Watanabe S., Tsuda Y. (2022). Olivine compositional variation of asteroid Ryugu samples: Possible precursors of Ryugu's parent asteroid. *53rd Lunar and Planetary Science Conference*, held March 7-11, 2022 in The Woodlands, Texas. LPI Contribution No. 1935.

Miyahara M., Yamaguchi A., Saitoh M., Fukimoto K., Sakai T., Ohfuji H., Tomioka N., Kodama Y. and Ohtani E. (2020) Systematic investigations of high-pressure polymorphs in shocked ordinary chondrites. *Meteoritics & Planetary Science* 55:2619-2651.

Morin G. L. F., Marrocchi Y., Villeneuve J., Jacquet E. (2022). $^{16}$O-rich anhydrous silicates in CI chondrites: Implications for the nature and dynamics of dust in the solar accretion disk. *Geochimica et Cosmochimica Acta* 332:203-219.

Nakamura T. (2005) Post-hydration thermal metamorphism of carbonaceous chondrites. *Journal of Mineralogical and Petrological Sciences* 100:260-272.

Nakato A., Nakamura T., Kitajima F., Noguchi T. (2008). Evaluation of dehydration mechanism during heating of hydrous asteroids based on mineralogical and chemical analysis of naturally and experimentally heated CM chondrites. *Earth Planet Space* 60 :855-864.

Noonan A. F. (1975). The Clovis (no. 1), New Mexico, Meteorite and Ca, AL and Ti-Rich Inclusions in Ordinary Chondrites. *Meteoritics* 10:51-59.

Noonan A. F., Fredriksson K., Jarosewich E., Brenner P. (1976). Mineralogy and bulk, chondrule, size-fraction chemistry of the Dhajala India chondrite. *Meteoritics* 11:340-343.

Pinto G. A., Marrocchi Y., Morbidelli A., Charnoz S., Varela M. E., Soto K., Martinez R., Olivares F. (2019). Constraints on Planetesimal Accretion Inferred from Particle-size Distribution in CO Chondrites. *The Astrophysical Journal Letters* 917:L25.

Prestgard T., Beck P., Bonal L., Eschrig J., Gattacceca J., Sonzogni C., Krämer-Ruggiu (2023). The parent bodies of CR chondrites and their secondary history. *Meteoritics and Planetary Science*.

Quirico E., Raynal P. I., Bourot-Denise M. (2003). Metamorphic grade of organic matter in six unequilibrated ordinary chondrites. *Meteoritics and Planetary Science* 38:795-811.

Quirico E., Bourot-Denise M., Robin C., Montagnac G., Beck P. (2011). A reappraisal of the metamorphic history of EH3 and EL3 enstatite chondrites. *Geochimica et Cosmochimica Acta* 75:3088-3102.

Regnault M., Marrocchi Y., Piralla M., Villeneuve J., Batanova V., Schnuriger N., Jacquet E. (2022). Oxygen isotope systematics of chondrules in Rumuruti chondrites: Formation conditions and genetic link with ordinary chondrites. *Meteoritics and Planetary Science* 57:122-135.

Righter K., Jakubek R. S., Fries M. D., Schutt J., Pando K., Harrington R. (2022). Assessment of petrologic subtypes, subgroups, and pairing within CV chondrites in the US Antarctic meteorite collection. *Meteoritics and Planetary Science* 58:25-40.

Righter K., Alexander C. M. O'D., Foustoukos D. I., Eckart L. M., Mertens C. A. K., Busemann H., Maden C., Schutt J., Satterwhite C. E., Harvey R. P., Pando K., Karner J. (2024). Pairing relations within CO3 chondrites recovered at the Dominion Range and Miller Range, Transantarctic mountains: Constraints from chondrule olivines, noble gas, and H, C, N bulk and isotopic compositions. *Meteoritics and Planetary Science* 59:1258-1276.

Roduit, N. (2020). JMicroVision: Image analysis toolbox for measuring and quantifying components of high-definition images. Version 1.3.2. https://jmicrovision.github.io.

Rubin A. E., Trigo-Rodriguez J. M., Huber H., Wasson J. T. (2007). Progressive aqueous alteration of CM carbonaceous chondrites. *Geochimica et Cosmochimica Acta* 71:2361-2382.

Rubin A. E. (2015). An American on Paris: Extent of aqueous alteration of a CM chondrite and the petrography of its refractory and amoeboid olivine inclusions. *Meteoritics and Planetary Science*, 50:1595-1612.

Rubin A E. and Li Y. (2019). Formation and destruction of magnetite in CO3 chondrites and other chondrite groups. *Geochemistry* 79:125528.

Russell S. S., Suttle M. D., King A. J. (2021). Abundance and importance of petrological type 1 chondritic material. *Meteoritics and Planetary Science*, 57:277-301.

Ruzicka A. (1990). Deformation and thermal histories of chondrules in the Chainpur (LL3.4) chondrite. *Meteoritics* 25:101-113.

Sakamoto N., Seto Y., Itoh S., Kuramoto K., Fujino K., Nagashima K., Krot A. N., Yurimoto H. (2007). Remnants of the Early Solar System Water Enriched in Heavy Oxygen Isotopes. Science 317:231.

Schrader D. L., Connolly H C. Jr, Lauretta D. S., Zega T. J., Davidson J., Domanik K. J. (2015). The formation and alteration of the Renazzo-like carbonaceous chondrites III: Toward understanding the genesis of ferromagnesian chondrules. *Meteoritics and Planetary Science* 50:15-50.

Schrader D. L., Davidson J. (2017). CM and CO chondrites: A common parent body or asteroidal neighbors? Insights from chondrule silicates. *Geochimica et Cosmochimica Acta* 214:157-171.

Schwinger S., Dohmen R., Schertl H.-P. (2016). A combined diffusion and thermal modeling approach to determine peak temperatures of thermal metamorphism experienced by meteorites. *Geochimica et Cosmochimica Acta* 191:255-276.

Scott E. R. D. (1984). Classification, Metamorphism, and Brecciation of Type 3 Chondrites from Antarctica. *Smithsonian Contributions to the Earth Sciences* 26:73-94.

Scott E. R. D., Taylor G. J. (1983). Chondrules and Other Components in C, O, and E Chondrites: Similarities in Their Properties and Origins. *Journal of Geophysical Research* 88:B275-B286.

Scott E. R. D., Jones R. H. (1990). Disentangling nebular and asteroidal features of CO3 carbonaceous chondrites. *Geochimica et Cosmochimica Acta* 54:2485-2502.

Scott E. R. D., Jones R. H., Rubin A. E. (1994). Classification, metamorphic history, and pre-metamorphic composition of chondrules. *Geochimica et Cosmochimica Acta* 58:1203-1209.

Sears D. W., Grossman J. N., Melcher C. L., Ross L. M., Mills A. A. (1980). Measuring metamorphic history in unequilibrated ordinary chondrites. *Nature* 287:791-795.

Sears D. W., Grossman J. N., Melcher C. L. (1982). Chemical and physical studies of type 3 chondrites—I: Metamorphism related studies of Antarctic and other type 3 ordinary chondrites. *Geochimica et Cosmochimica Acta* 46:2471-2481.

Sears D. W. G., Batchelor J. D., Lu J., Keck B. D. (1991a). Metamorphism of CO and CO-like chondrites and comparisons with type 3 ordinary chondrites. *Antarctic Meteorite Research* 4:319-343.

Sears D. W. G., Hasan F. A., Batchelor J. D., Lu J. (1991b). Chemical and Physical Studies of Type 3 Chondrites—XI: Metamorphism, Pairing, and Brecciation of Ordinary Chondrites. *Proceedings of Lunar and Planetary Science* 21:493-512.

Sears D. W. (2016). The CO chondrites: Major recent Antarctic finds, their thermal and radiation history, and describing the metamorphic history of members of the class. *Geochimica et Cosmochimica Acta* 188:106-124.

Skirius C., Steele I. M., Smith J. V. (1986). Belgica-7904: A new carbonaceous chondrite from Antarctica: minor-element chemistry of olivine. *Proceedings of the Tenth Symposium on Antarctic Meteorites* 243-258.

Stöffler D., Keil K., Scott E. R. D. (1991). Shock metamorphism of ordinary chondrites. *Geochimica et Cosmochimica Acta* 55:3845-3867.

Stöffler D., Hamann C., Metzler K. (2018). Shock metamorphism of planetary silicate rocks and sediments: proposal for an updated classification system. Meteoritics and Planetary Science 53:5-49.

Suttle, M. D.; Greshake, A.; King, A. J.; Schofield, P. F.; Tomkins, A. and Russell, S. S. (2021a). The alteration history of the CY chondrites, investigated through analysis of a new member: Dhofar 1988 *Geochimica et Cosmochimica Acta* 295:286-309.

Suttle M. D., King A. J., Schofield P. F., Bates H., Russell S. S. (2021b). The aqueous alteration of CM chondrites, a review. *Geochimica et Cosmochimica Acta* 299:219-256.

Tomeoka K. (1990). Mineralogy and petrology of Belgica-7904: A new kind of carbonaceous chondrite from Antarctica. *Antarctic Meteorite Research* 3:40-54.

Tonui E., Zolensky M., Hiroi T., Nakamura T., Lipschutz M. E., Wang M.-S., Okudaira K. (2014). Petrographic, chemical and spectroscopic evidence for thermal metamorphism in carbonaceous chondrites I: CI and CM chondrites. *Geochimica et Cosmochimica Acta* 126:284-306.

Van Schmus W. R., Wood J. A. (1967). A chemical-petrologic classification for the chondritic meteorites. *Geochimica et Cosmochimica Acta* 31:747-765.

Van Schmus, W.R. (1969). Mineralogy, Petrology, and Classification of Types 3 and 4 Carbonaceous Chondrites. In: Millman, P.M. (eds) Meteorite Research. Astrophysics and Space Science Library, vol 12. Springer, Dordrecht.

Van Schmus W. R., Hayes J. M. (1972). Chemical and petrographic correlations among carbonaceous chondrites. *Geochimica et Cosmochimica Acta* 38:47-64.

Velbel M. A., Tonui E. K., Zolensky M. E. (2012). Replacement of olivine by serpentine in the carbonaceous chondrite Nogoya (CM2). *Geochimica et Cosmochimica Acta* 87:117-135.

Velbel M. A., Tonui E. K., Zolensky M. E. (2015). Replacement of olivine by serpentine in the Queen Alexandra Range 93005 carbonaceous chondrite (CM2): Reactant-product compositional relations, and isovolumetric constraints on reaction stoichiometry and elemental mobility during aqueous alteration. *Geochimica et Cosmochimica Acta* 148:402-425.

Weisberg M.K., Prinz M., Clayton R.N., Mayeda T.K. (1993). The CR (Renazzo-type) carbonaceous chondrite group and its implications. *Geochimica et Cosmochimica Acta* 57:1567-1586.

Weisberg M. K., Prinz M., Clayton R. N., Mayeda T. K., Grady M. M., Pillinger C. T. (1995). The CR chondrite clan. Proceedings of the NIPR Symposium on Antarctic Meteorites 8:11-32.

Weisberg, M. K., Prinz, M., Clayton, R. N. & Mayeda, T. K. (1997) CV3 chondrites: three subgroups, not two. Meteoritics and Planetary Science Supplements 32 :138–139.

Weisberg M.K. and Huber H. (2007) The GRO 95577 CR1 chondrite and hydration of the CR parent body. *Meteoritics & Planetary Science* 42:1495-1503.

Weisberg M. K., McCoy T. J., Krot A. N. (2006). Systematics and Evaluation of Meteorite Classification. In *Meteorites and the Early Solar System II*, p 19-52, eds. D. S. Lauretta and H. Y. McSween Jr, Cambridge University Press.

Weyrauch M., Horstmann M., Bischoff A. (2018). Chemical variations of sulfides and metal in enstatite chondrites—Introduction of a new classification scheme. *Meteoritics and Planetary Science* 53:394-415.

Wick M. J., Jones R. H. (2012). Formation conditions of plagioclase-bearing type I chondrules in CO chondrites: A study of natural samples and experimental analogs. *Geochimica et Cosmochimica Acta* 98:140-159.

Wight M. J., Bland P., Wright I. P., Franchi I. A., Pillinger C. T. (1994). Equilibration Temperature of "Reduced Ordinary Chondrites". *Meteoritics* 29:551-552.

Wiik H. B. (1956). The chemical composition of some stone meteorites. *Geochimica et Cosmochimica Acta* 9:279-289.

Wlotzka F. (1993). A Weathering Scale for the Ordinary Chondrites. *Meteoritics* 28:460-460.

Wood J. A. (1967). Olivine and pyroxene compositions in type II carbonaceous chondrites. *Geochimica et Cosmochimica Acta* 31:2095-2108.

Wood B. J. and Blundy J. D. (2014). Trace element partitioning: The influences of ionic radius, cation charge, pressure, and temperature. In Meteorites and cosmochemical processes, edited by Davis A. M. Treatise on Geochemistry. Amsterdam: Elsevier. pp. 421–448.

Zanda B., Hewins R. H., Bourot-Denise M., Bland P. A., Albarède F. (2006). Formation of solar nebula reservoirs by mixing chondritic components. *Earth and Planetary Science Letters* 248:650-660.

Zhang Y., Benoit P. H., Sears D. W. G. (1995). The classification and complex thermal history of the enstatite chondrites. *Journal of Geophysical Research* 100:9417-9438.

Zhang Y., Huang S., Schneider D., Benoit P. H., DeHart J. M., Lofgren G. E., Sears D. W. G. (1996). Pyroxene structures, cathodoluminescence and the thermal history of the enstatite chondrites. *Meteoritics and Planetary Science* 31:87-96.

Zolensky M. E., Mittlefehldt D. W., Lipschutz M. E., Wang M.-S., Clayton R. N., Mayeda T. K., Grady M. M., Pillinger C., Barber D. (1997). CM chondrites exhibit the complete petrologic range from type 2 to 1. *Geochimica et Cosmochimica Acta* 61:5099-5115.